\documentclass[runningheads,a4paper]{llncs}

\usepackage{amssymb}
\usepackage{graphicx}

\usepackage{mathtools}
\usepackage{mathrsfs}
\usepackage{float}
\usepackage{stmaryrd}
\usepackage{units}
\usepackage{dsfont}
\usepackage{pgfplots}
\usepackage{framed}
\usepackage{listings}
\usepackage{cite}

\newcommand{\mchain}{\mathscr{M}} % The symbol for Markov Chain tuple
\newcommand{\msspace}{S} % The symbol for state space
\newcommand{\mmbl}{\Sigma} % The symbol for measurable sets of states
\newcommand{\mker}{\tau} % The symbol for transition kernel
\newcommand{\mlabel}{L} % The symbol for the labelling function
\newcommand{\mtuple}{{\left( \msspace, \mmbl, \mker, \mlabel \right)}}
\newcommand{\mvar}[2]{\mchain^{#1}_{#2}} % The symbol for chain as a random variable, with initial dist #1, at time #2
\newcommand{\prob}[3]{P_{#1}\!\left(#2,#3\right)} % The probability of following length #1 trace #3 from state #2

\usepackage{url}

\begin{document}

\mainmatter  % start of an individual contribution

\title{On the Relationship between Bisimulation and Trace Equivalence in an Approximate Probabilistic Context (Extended Version)}

\titlerunning{Bisimulation vs Trace Equivalence in an Approximate Probabilistic Context}

\author{Gaoang Bian \inst{1}\inst{2}
\and Alessandro Abate \inst{2}}

\institute{Google Inc, United States 
\and Dept. of Computer Science, University of Oxford, UK}

\authorrunning{G. Bian and A. Abate}

\toctitle{Bisimulation vs Trace Equivalence in an Approximate Probabilistic Context}
\tocauthor{G. Bian and A. Abate}
\maketitle

\begin{abstract}
This work introduces a notion of approximate probabilistic trace equivalence for labelled Markov chains, 
and relates this new concept to the known notion of approximate probabilistic bisimulation. 
In particular this work shows that the latter notion induces a tight upper bound on the approximation between finite-horizon traces, 
as expressed by a total variation distance. 
As such, 
this work extends corresponding results for exact notions and analogous results for non-probabilistic models. 
This bound can be employed to relate the closeness in satisfaction probabilities over bounded linear-time properties, 
and allows for probabilistic model checking of concrete models via abstractions.  
The contribution focuses on both finite-state and uncountable-state labelled Markov chains, and claims two main applications:   
firstly, it allows an upper bound on the trace distance to be decided for finite state systems; 
secondly, it can be used to synthesise discrete approximations to continuous-state models with arbitrary precision.
\end{abstract}

\section{Introduction}

Often in formal verification 
%and a huge variety of other fields,
one is interested in approximations of 
%Markov 
concrete models. 
Models are often built from experimental data that are themselves approximate, 
and taking approximations can reduce the size and complexity of the state space. 
%One particular use case that we will focus on is the automated model checking\cite{KNP11, KZHHJ09} of continuous state models by approximating them with discrete models. 
Markov models in particular can be defined either syntactically as a transition structure (with states and matrices), 
or semantically as a random process whose trajectory satisfy the 
%time homogeneous 
Markov property. 
Each representation gives rise to its own notions of approximation\cite{A13}: ``the transition matrices have similar numbers and/or structure'' vs ``the trajectories have similar probability distributions'', respectively. 
While the syntactic representation is used for computations and model checking with concrete numbers, often one is interested in results in terms of the semantics, e.g. ``what is the probability of reaching a failure state within 100 steps''. This gives practical value to studying how approximations in terms of transition matrices translate into approximations in terms of traces of the random process. 

In this paper we build on the notion of $\varepsilon$-approximate probabilistic bisimulation, introduced in \cite{DLT08} as a natural extension to exact probabilistic bisimulation \cite{DEP02}. 
There, the notion of $\varepsilon$-approximate probabilistic bisimulation (or just $\varepsilon$-bisimulation) is defined in terms of the transition structure,
and given $\varepsilon$  
%\alex{[are we sure? I thought this work just computed an $\varepsilon$-bisimulation]}
%\colin{[I'm quite sure it is the maximal. 
%Note that $\sim_\varepsilon$ is defined in that paper to relate states that are $\varepsilon$-bisimilar (under any $\varepsilon$-bisimulation relation),
%and this is exactly the maximal $\varepsilon$-bisimulation relation.
%Plus it says ``one could be interested in computing
%efficiently the largest relation''
%]}
the maximal $\varepsilon$-bisimulation relation can be computed for finite state systems with $n$ states in $\mathcal O(n^7)$ time \cite{DLT08}.  

\begin{figure}[t]
\label{expl:branching time too strong}

\centering
%trim=left bottom right top
\includegraphics[clip, width=80mm, trim=5cm 19.85cm 5.5cm 4.6cm]{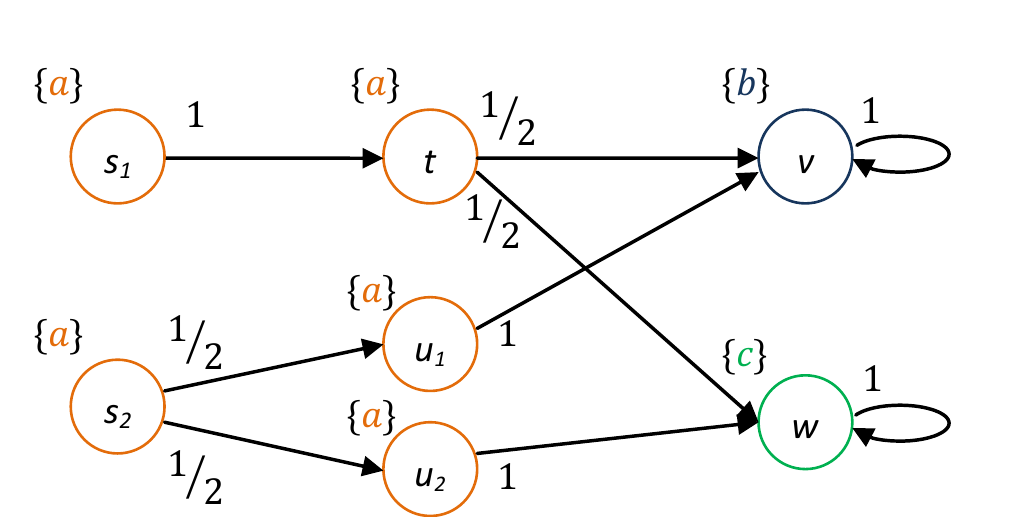}

\caption{
\textbf{Branching vs Linear Time Behaviour.}
In the Labelled Markov Chain below (cf. Section \ref{sect:preliminaries} for the LMC model), 
the states $s_1, s_2$ both emit traces $ \langle \{a\}, \{a\}, \{b\} \rangle $ and $ \langle \{a\}, \{a\}, \{c\} \rangle $ with probability $0.5$ each, and hence $s_1, s_2$ have the same linear time behaviour.
However, $s_1, s_2$ have different branching behaviour, since exclusively 
$s_1$ satisfies the PCTL formula 
$ \mathrm{P}_{=1}\left[ \; \mathrm{X} \; \mathrm{P}_{=0.5}\left[ \; \mathrm{X} \; b \; \right] \; \right] $. 
%, but $s_2$ does not. 
Conversely, only $s_2$ satisfies
$ \mathrm{P}_{=0.5}\left[ \; \mathrm{X} \; \mathrm{P}_{=1}\left[ \; \mathrm{X} \; b \; \right] \; \right] $. 
}
\end{figure}

It is on the other hand of interest to explore what $\varepsilon$-bisimulation means in terms of trajectories. While $\varepsilon$-bisimulation does have characterizations (on countable state spaces) in terms of logics and games \cite{DEP02}, this logic is branching in nature, and does not directly relate to the trajectory of the model as it leaks information about the state space (similarly to the difference between CTL and LTL), as illustrated in Figure \ref{expl:branching time too strong}.

In this paper, we investigate what $\varepsilon$-approximate probabilistic bisimulation means in terms of trajectories. We will prove that for Labelled Markov Chains (over potentially uncountable state spaces), $\varepsilon$-bisimulation between two states places the tight upper bound of $1-(1-\varepsilon)^k$ (which is $\leq k\varepsilon$) on the total variation \cite{GS02} between the distributions of length $k+1$ traces starting from those states, for all $k \in \mathbb N$.
We will formulate these bounds by introducing the notion of $f(k)$-trace equivalence.
%That is, the probability assigned to any set of length $k+1$ traces differs by at most $1-(1-\varepsilon)^k$ between two states that are $\varepsilon$-bisimilar. We will encapsulate these bounds as $\varepsilon$-probabilistic trace equivalence (or just $\varepsilon$-trace equivalence), so that $\varepsilon$-trace equivalence induces a tight bound on the total variation of the distribution of length $k$ traces (for each $k$) guaranteed by $\varepsilon$-bisimulation.
As such, we extend the well known result that bisimulation implies trace equivalence in non-probabilistic systems to the context of approximate and probabilistic models (the exact probabilistic case having been considered in \cite{CPP}).

One direct repercussion of our result is that it provides a method to efficiently bound the total variation of length $k$ traces from two finite-state LMCs (or two states in an LMC), 
since the aforementioned result in \cite{DLT08} can be used to decide or to compute 
%\alex{[rather, to compute - indeed, I think $\varepsilon$ is computed by the algorithm, not given a-priori]}
%\colin{[It says
%``Given e > 0. one could be interested in computing
%efficiently the largest relation of e-simulation
%...'',
%and then later on,
%``we can in a straightforward way, using a dichotomic approach, compute the distance $d(s,t)$ between two states ... up to precision $\delta$ in time $\mathcal O (\left|S\right|^7 \cdot ln(\delta))$.
%''
%]}
$\varepsilon$-bisimulation between two states in polynomial time. 
We will also apply our results to the quantitative verification of continuous-state Markov models \cite{AKLP10, AKNP14, SA13}, 
improving on the current class of properties approximated and the corresponding approximation errors. 

\smallskip

\noindent\textbf{Related Work. }
Literature on approximations of (finite-state) Markov models can be distinguished into two main branches: 
one focusing on one-step similarity, 
the other dealing with trace distances. 
One-step similarity can be studied via the notion of probabilistic bisimulation, 
introduced in the context of finite-state models by \cite{larsen}, 
and related to lumpability in \cite{SD06}. 
%Contrary to this exact notion,  
\cite{DLT08} discusses a notion of approximate bisimulation, 
related to quasi-lumpability conditions in \cite{BP94, s96}.
%\alex{Comment on \cite{GT13}}
From the perspective of process algebra, 
\cite{GT13} studies operators on probabilistic transition systems that preserve the approximate bisimulation distance.  
% The next sentence was moved to later.
%\cite{PLS13, WTM12} study CTL and LTL properties of LMCs with uncertainties in their transition probabilities. 
%\footnote{\colin{deleted: which is related to the approach in \cite{IAK12}.}}
The work in \cite{DLT08,Segala1995} is seminal in introducing notions of (exact) probabilistic simulation, much extended in subsequent literature.  

On the other hand, there are a few papers studying the total variation distance over traces. 
\cite{CK14} presents an algorithm for approximating the total variation of infinite traces of labelled Markov systems and prove the problem of deciding whether it exceeds a given threshold to be NP-hard. 
\cite{CFW12} shows that the undiscounted bisimilarity pseudometric is a (non-tight) upper bound on the total variation of infinite traces
(like $\varepsilon$-bisimulation, the bisimilarity pseudometric is defined on the syntax of the model
and there are efficient algorithms for computing it \cite{CFW12, CPP12, TB16}).
The contribution in this paper, 
focusing on finite traces rather than infinite traces, 
is that the total variation of finite traces is much less conservative, 
and moreover allows manipulating models 
%to be synthesised with 
under specific error bounds on length $k$ traces, as we will show in Section \ref{sect: application to model checking}.

\cite{IAK12} studies notions based on the total variation of finite and infinite traces: 
employing a different notion of $\varepsilon$-bisimulation than ours, it proves error bounds on trace distances, 
which however depend on additional properties of the structure of the transition kernel
(as shown in Section \ref{sect: alternative notions} the error bound on reachability probabilities could go to $1$ in two steps, for any $\varepsilon > 0$).
Finite abstractions of continuous-state Markov models can be synthesised by notions that are variations of the 
$\varepsilon$-bisimulation in this work \cite{A13,AKNP14,SA13}.  
%\cite{PLS13, WTM12} study CTL and LTL properties of LMCs with uncertainties in their transition probabilities. 
Tangential to our work, \cite{DHKP16} shows that the total variation of finite traces can be statically estimated via repeated observations. 
\cite{W10} investigates ways of compressing Hidden Markov Models by searching for a smaller model that minimises the total variation of length-$k$ traces of the two models.

%In a non-probabilistic context \ldots : 
%The notion of $\delta$-approximate bisimulation\cite{GP05} is based on distances between states (based on an existing metric over the state space) rather than differences in transition probabilities. 

%There are several notions of approximation on Markov systems based on their traces (e.g. \cite{DGJP04, LZ06}).

%----

% Consider also refrencing these:

% Estimate discounted trace equivalence (of finite or infinite traces) through reinforcement learning\cite{LZ06}. Not clear how trace equivalence is approximated (total variation or something else).

% Markov Chain Lumpability on Fuzzy Partitions

% LZ06 (consider expanding): Total variation of infinite traces for Semi-Markov Chains (continuous time). Proves that it is related to something else: https://scholar.google.com/scholar?cluster=9884782410329577645&hl=en&as_sdt=0,5&sciodt=0,5

\smallskip 

\noindent\textbf{Structure of this article. }
In Section \ref{sect:preliminaries}, we introduce the reference model (labelled Markov chains -- LMC -- over general state spaces) and provide a definition of $\varepsilon$-bisimulation for LMCs. 
In Section \ref{sect: approx trace equiv}, we introduce the notion of
%approximate trace equivalence,
approximate probabilistic trace equivalence (and the derived notion of probabilistic trace distance), 
and discuss how it relates to bounded linear time properties, 
and to the notion of distinguishability. 
In Section \ref{sect: approx bisim implies approx trace equiv}, we present the main result: 
we will derive a tight upper bound on the probabilistic trace distance between $\varepsilon$-bisimilar states. 
In Section \ref{sect: application to model checking}, we show how these results can be used to approximately model check continuous state systems, 
and 
%relate the outcomes to existing work.  
%improving on the current class of properties approximated and the approximation errors.
%
Section \ref{sect:application case study} discusses a case study. 
%and Section \ref{sect: alternative notions} reviews an alternative definition of approximate probabilistic bisimulation in the literature,
%demonstrating that it does not by itself guarantee closeness of bounded reachability probabilities.
In Section \ref{sect: alternative notions}, we discuss an alternative notion of approximate probabilistic bisimulation that appears in literature and show that it cannot be used to effectively bound probabilistic trace distance.
The Appendix contains the lengthier proofs of some the statements in this work and details on the implementation of the Case Study.

\section{Preliminaries}
\label{sect:preliminaries}

%In this section, we first provide a definition for labelled Markov Chains and discuss it's relation to other models used in related literature. Then we briefly review the existing work on exact probabilistic bisimulation and $\varepsilon$-bisimulation that we build on. Finally, we extend $\varepsilon$-bisimulation to uncountable models and show that some basic properties are preserved.

\subsection{Labelled Markov Chains}

We will work with discrete-time Labelled Markov Chains (LMCs) over general state spaces.
Known definitions of countable- or finite-state LMCs 
%\cite{IAK12}[\colin{Wrong citation?}]
represent special instances of the general models we introduce next. 

\begin{definition}[LMC syntax]
\label{def:LMC}
A Labelled Markov Chain (LMC) is a structure
$
\mchain = \mtuple
$ where:
\begin{itemize}
\item $\msspace$ is a (potentially uncountable) set of states. 
\item $\mmbl \subseteq \mathcal P(\msspace)$ is a $\mmbl$-algebra over $\msspace$ representing the set of measurable subsets of $\msspace$. 
\item $\mker: \msspace \times \mmbl \to [0,1]$ is a transition kernel. 
That is, for all $s\in\msspace$, $\mker(s,\cdot)$ is a probability measure on the measure space $(\msspace, \mmbl)$, 
and for all $A \in \mmbl$ we require $\mker(\cdot,A)$
to be $\mmbl$-measurable. 
\item $\mlabel: \msspace \to \mathcal O$ labels each state $s \in S$ with a subset of atomic propositions from $\mathrm{AP}$, where $\mathcal O = 2^\mathrm{AP}$. $L$ is required to be $\mmbl$-measurable, and we will assume $\mathrm{AP}$ to be finite. %\qed
\end{itemize}
\end{definition}
$\mlabel(s)$ captures all the observable information at state $s \in \msspace$: 
this drives our notion relating pairs of states, 
and we characterise properties over the codomain of this function. 

\begin{definition}[LMC semantics]
\label{def:LMC trajectory}
Let $\mchain = \mtuple$ be a LMC. 
Given an initial distribution $p_0$ over $\msspace$, 
the state of $\mchain$ at time $k$
%$t \in \mathbb N$ 
is a random variable $\mvar{p_0}{k}$ 
over $\msspace$, such that 

\begin{equation*}
\begin{split}
&\mathbb P \left[ \mvar{p_0}{0} \in A_0 \right] = p_0(A_0), 
\\&\mathbb P \left[ \mvar{p_0}{0} \in A_0, \cdots, \mvar{p_0}{k} \in A_k \right]
=\int_{y_0 \in A_0} p_0(\mathrm{d}y_0)
\\&\indent\indent \cdotp\; \int_{y_1 \in A_1} \mker(y_0,\mathrm{d}y_1)
\cdots \int_{y_{k-1} \in A_{k-1}} \mker(y_{k-2},\mathrm{d}y_{k-1})
\;\cdotp\; \mker(y_{k-1}, A_k), 
\end{split}
\end{equation*}
for all $k\in\mathbb N \backslash \{0\}, A_k \in \mmbl$, where of course $\mker(y_{k-1}, A_k) = \int_{y_{k} \in A_{k}} \mker(y_{k-1},\mathrm{d}y_{k})$. 
%\qed 
\end{definition}  
%Appendix \ref{sect:models_in_related_work} discusses the relationship with models in the literature. 
%\alex{[next statement is unclear]}
%We consider two states $s_1, s_2 \in \msspace$ to look the same at time $t \in \mathbb N$ iff $\mlabel(s_1) = \mlabel(s_2)$, and as we shall see this drives the notions of $\varepsilon$-bisimulation and $\varepsilon$-trace equivalence.

%\alex{[Im moving this text back here - thus leaving in the appendix almost exclusively extended proofs and details of case studies.]}
\noindent\textbf{Models in related work.} 
%\label{sect:models_in_related_work}
A body of related literature works with labelled MDPs, which are more general models allowing a non-deterministic choice $u \in \mathcal U$ (for some finite $\mathcal U$) of the transition kernel $\mker_u$ at each step. 
%Labelled MDPs evolve under an external ``policy'' $\sigma : \{ \langle \alpha_1, \cdots, \alpha_k \rangle \mid \alpha_i \in 2^{\mathrm{AP}}\} \to \mathrm{dist}_\mathcal U$, which maps the sequence of past observations of the process to a probability distribution over $\mathcal U$, from which the next action $u \in \mathcal U$ is drawn.
This choice is made by a ``policy'' that probabilistically selects $u$ based on past observations of the process.
% policies must be label-measurable
Whilst we will ignore non-determinism and work with LMCs for simplicity, our results can be adapted to 
%valid over
labelled MDPs by quantifying over all policies, or over all choices $u$ for properties like $\varepsilon$-bisimulation, in order to remove the non-determinism. 
The seminal work on bisimulation and $\varepsilon$-bisimulation dealt with models known as LMPs \cite{DEP02}. 
LMPs allow for non-determinism (like labelled MDPs) but their states are unlabelled and at each step they have a probability of halting. For the study of bisimulation in this work, LMPs can be considered as a simplification of labelled MDPs to the case $\mathcal O = \{\emptyset,\{\mathrm{halted}\}\}$. 
%However, the fact that the halted state is not considered part of the state space allows for more interesting concepts of Simulation and two-way Simulation (as discussed in \cite[Section 2.1]{DLT08}). 

\subsection{Exact and Approximate Probabilistic Bisimulations}
\label{subsect: def approx bisim}

The notion of approximate probabilistic bisimulation (in this work just $\varepsilon$-bisimulation) is a structural notion of closeness, 
based on the stronger notion of exact probabilistic bisimulation \cite{DEP02}. 
We discuss both next. 
Considering a binary relation $R$ over set $X$, 
we say that a subset $\tilde S \subseteq X$ is $R$-closed if $\tilde S$ contains its own image under $R$. That is, if $R(\tilde S) \coloneqq \{ y \in X \mid x \in \tilde S, \; x R\, y\} \subseteq \tilde S$.

\begin{definition}[Exact probabilistic bisimulation]
\label{def:Exact probabilistic bisimulation}
Let $\mchain = \mtuple$ be a LMC. 
%For $\varepsilon \in [0,1]$, 
An equivalence relation $R \subseteq \msspace \times \msspace$ over the state space is an exact probabilistic bisimulation relation if 
\begin{align*}
  & \forall (s_1, s_2) \in R, \indent\text{we have that } \mlabel(s_1) = \mlabel(s_2),
\\& \forall (s_1, s_2) \in R, \; \forall \tilde T \in \mmbl \text{ s.t. $\tilde T$ is $R$-closed},
    \indent\text{we have that } \mker(s_1,\tilde T) = \mker(s_2,\tilde T). 
\end{align*}
A pair of states $s_1, s_2 \in \msspace$ are said to be (exactly probabilistically) bisimilar if there exists an exact probabilistic bisimulation relation $R$ such that $s_1 R\, s_2$. 
%\qed
\end{definition}
Note that since $R$ is an equivalence relation, $R$-closed sets are exactly the unions of whole equivalence classes.

Next, we adapt the notion of $\varepsilon$-bisimulation (as discussed in \cite{DLT08} for LMPs over countable state spaces) to LMCs over general spaces.
%We have introduced the above notion over general LMCs: next,   
%we adapt the notion of $\varepsilon$-bisimulation 
%(as discussed in \cite{DLT08} for LMPs)
%to LMCs over countable state spaces, as follows.  

\begin{definition}[$\varepsilon$-bisimulation]
\label{defi:approx bisim countable} 
Let $\mchain = \mtuple$ be a LMC. For $\varepsilon \in [0,1]$, a symmetric binary relation $R_\varepsilon \subseteq \msspace \times \msspace$ over the state space is an $\varepsilon$-approximate probabilistic bisimulation relation (or just $\varepsilon$-bisimulation relation) if 
\begin{align}
  &
  \forall T \in \mmbl, \indent\text{we have } R_\varepsilon(T) \in \mmbl,
\label{eqn:R is mbl}
\\& \forall (s_1, s_2) \in R_\varepsilon, \indent\text{we have } \mlabel(s_1) = \mlabel(s_2),
\\& \forall (s_1, s_2) \in R_\varepsilon, \; \forall T \in \mmbl,
    \indent\text{we have } \mker(s_2,R_\varepsilon (T)) \geq \mker(s_1,T) - \varepsilon. 
\label{eqn:approx bisim countable}
\end{align}
%\alex{[what if the choice of $\varepsilon$ makes the RHS negative? I attempt an obvious fix, to be propagated below.]} 
%\colin{[TODO: discuss in next VC]}
Two states $s_1, s_2 \in \msspace$ are said to be $\varepsilon$-bisimilar
if there exists an $\varepsilon$-bisimulation relation $R_\varepsilon$ such that $s_1 R_\varepsilon s_2$. 
%\qed
\end{definition}

The condition raised in \eqref{eqn:approx bisim countable} could be understood intuitively as 
``for any move that $s_1$ can take (say, into set $T$), 
$s_2$ can match it with higher likelihood over the corresponding set $R_\varepsilon (T)$, up to $\varepsilon$ tolerance.''
Notice that (\ref{eqn:R is mbl}) is not a necessary requirement for countable state models, but for uncountable state models it is needed to ensure that $R_\varepsilon (T)$ is measurable and $\mker(s_2,R_\varepsilon (T))$ is defined in (\ref{eqn:approx bisim countable}). 

\cite{DLT08} showed that in countable state spaces, $0$-approximate probabilistic bisimulation corresponds to exact probabilistic bisimulation. On uncountable state spaces, not every exact probabilistic bisimulation relation is a $0$-bisimulation relation because of the additional measurably requirement, but we still have that $0$-bisimulation implies exact probabilistic bisimulation.

\begin{theorem}\label{thm:0bisim}
Let $\mchain = \mtuple$ be a LMC, and let $s_1,s_2 \in \msspace$. If $s_1,s_2$ are $0$-bisimilar, then they are (exactly, probabilistically) bisimilar.  
\end{theorem}

Although above $s_1, s_2$ are required to belong to the state space of a given LMC, 
the notions of exact- and $\varepsilon$-bisimulation can be extended to hold over pairs of LMCs by combining their state spaces, as follows.

%\alex{[I've merged two definition into the one below.]}
%
\begin{definition}[$\varepsilon$-bisimulation of pairs of LMCs]
\label{def:direct sum of LMCs}
Consider two LMCs $\mchain_1 = \left( \msspace_1, \mmbl_1, \mker_1, \mlabel_1 \right)$ and $\mchain_2 = \left( \msspace_2, \mmbl_2, \mker_2, \mlabel_2 \right)$.
Without loss of generality, 
assume that their state spaces $\msspace_1, \msspace_2$ are disjoint. 
The direct sum $\mchain_1 \oplus \mchain_2$ of $\mchain_1$ and $\mchain_2$ is the LMC formed by combining the state spaces of $\mchain_1$ and $\mchain_2$. 
Formally, $\mchain_1 \oplus \mchain_2 = \left( \msspace_1 \uplus \msspace_2, \sigma \left( \mmbl_1 \times \mmbl_2 \right), \mker_1 \oplus \mker_2, \mlabel_1 \uplus \mlabel_2 \right)$, 
where:
\begin{itemize}
\item $\msspace_1 \uplus \msspace_2$ is the union of $\msspace_1$ and $\msspace_2$ where we have assumed wlog (by relabelling if necessary) that $\msspace_1,\msspace_2$ are disjoint;
\item $\sigma \left( \mmbl_1 \times \mmbl_2 \right)$ is the smallest $\sigma$-algebra containing $\mmbl_1 \times \mmbl_2$;
\item $\mker_1 \oplus \mker_2 \left( s, T \right) \coloneqq \begin{cases}
\mker_1(s,\; T \cap \msspace_1) &\mbox{if } s \in \msspace_1 \\ 
\mker_2(s,\; T \cap \msspace_2) &\mbox{if } s \in \msspace_2 
\end{cases}$
for $s \in \msspace_1 \uplus \msspace_2$, $T \in \sigma \left( \mmbl_1 \times \mmbl_2 \right)$;
\item $\mlabel_1 \uplus \mlabel_2 (s) \coloneqq \begin{cases}
\mlabel_1(s) &\mbox{if } s \in \msspace_1 \\ 
\mlabel_2(s) &\mbox{if } s \in \msspace_2 
\end{cases}$
for $s \in \msspace_1 \uplus \msspace_2$.
%\\\indent\indent for $s \in \msspace_1 \uplus \msspace_2$. %\qed
\end{itemize} 
Let $s_1 \in \msspace_1$, $s_2 \in \msspace_2$. 
We say that $s_1, s_2$ are $\varepsilon$-bisimilar iff $s_1, s_2$ are $\varepsilon$-bisimilar as states in the direct sum LMC $\mchain_1 \oplus \mchain_2$. %\qed
\end{definition}

\noindent \textbf{Other Notions of $\varepsilon$-Bisimulation in Literature. }
There is an alternative, more direct, extension of exact probabilistic bisimulation in literature \cite{A13, AKNP14, IAK12}, which simply requires $| \mker(s_1,\tilde T) - \mker(s_2,\tilde T) | \leq \varepsilon$ instead of the conditions in Def. \ref{defi:approx bisim countable}.
However, this requirement alone is too weak to guarantee properties that we later discuss (cf. Section \ref{sect: alternative notions}).

\section{Approximate Probabilistic Trace Equivalence for LMCs}
\label{sect: approx trace equiv}

In this section we introduce the concept of approximate probabilistic trace equivalence
%/\alex{probabilistic trace Distance} 
(or just $f(k)$-trace equivalence) 
%/\alex{$\varepsilon$-trace Distance}) 
to represent closeness of observable linear time behaviour. 
Based on the likelihood over traces of a given LMC, 
this notion depends on its operational semantics (cf. Definition \ref{def:LMC trajectory}), 
rather than on the structure of its transition kernel (as in the case of approximate bisimulation). 
The notion can alternatively be thought of inducing a distance among traces, as elaborated below.  

\begin{definition}[Trace likelihood]
\label{def:ProbTrace}
Let $\mchain = \mtuple$ be an LMC, $s_0 \in \msspace$, and $k \in \mathbb N$.  
Let $\mathrm{TRACE}$ denote a set of traces (each of length $k+1$), 
taking values in time over $2^{\mathrm{AP}}$,
so that $\mathrm{TRACE} \subseteq \mathcal O^{k+1}$.
%namely 
%\begin{align*}
%&\mathrm{TRACE} = \{ \mathrm{trace}_1, \cdots, \mathrm{trace}_n \}, 
%\\&\indent \text{where\indent} \mathrm{trace}_i = \langle\; \alpha_{i\;0}, \cdots, \alpha_{i\;k} \;%\rangle
%\indent \text{and \indent} \alpha_{i\;j} \in \mathcal O = 2^\mathrm{AP}. 
%\end{align*}
\noindent
Denote with $\prob{k}{s_0}{\mathrm{TRACE}}$ the probability that the LMC $\mchain$, 
given an initial state $s_0$,  
generates any of the runs $\langle \alpha_0, \cdots, \alpha_k \rangle \in \mathrm{TRACE}$, 
namely 
\begin{align*}
&\prob{k}{s_0}{\mathrm{TRACE}}
 = \sum_{\substack{ \langle \alpha_0, \cdots, \alpha_k \rangle \\\in \mathrm{TRACE}}} \mathbb P \big[ \mvar{s_0}{0} \in \mlabel^{-1}(\{\alpha_{0}\}), \cdots,
\mvar{s_0}{k} \in \mlabel^{-1}(\{\alpha_{k}\}) \big], 
\end{align*}
where $\mvar{s_0}{t}$ is the state of $\mchain$ at step $t$, 
with a degenerate initial distribution $p_0$ that is concentrated on point $s_0$ (cf. Definition \ref{def:LMC trajectory}). %\qed 
\end{definition}
As intuitive, 
we consider traces of length $k+1$ (rather than of length $k$) because a length $k+1$ trace is produced by one initial state and precisely $k$ transitions. 
Notice that the set of sequences of states generating $\mathrm{TRACE}$ is measurable, 
being defined via a measurable map $L$ over a finite set of traces.

\begin{definition}[Total variation \cite{D04}]
Let $(Z,\mathcal G)$ be a measure %measurable
space where $\mathcal G$ is a $\sigma$-algebra over $Z$, and let $\mu_1, \mu_2$ be probability measures over $(Z,\mathcal G)$. The total variation between $\mu_1, \mu_2$ is
%\begin{align*}
$d_\mathrm{TV}(\mu_1,\mu_2) \coloneqq \sup_{A \in \mathcal G} \left| \mu_1(A) - \mu_2(A) \right|$.  
%\end{align*}
%\qed
\end{definition}

%\alex{
%\begin{definition}[$\varepsilon$-trace \alex{Distance}]
%Let ${\mchain = \mtuple}$ be a LMC. 
%Consider a non-decreasing function $\varepsilon: \mathbb N \rightarrow [0,1]$. 
%We say that states $s_1,s_2 \in \msspace$ have an $\varepsilon$-probabilistic trace Distance (or are just ``$\varepsilon$-trace Distanct'') if for all $k \in \mathbb N$,
%\begin{align*}
%d_\mathrm{TV}\big(\prob{k}{s_1}{\cdot},\prob{k}{s_2}{\cdot}\big) \leq \varepsilon (k),
%\end{align*}
%or equivalently if for all $k \in \mathbb N, \,\mathrm{TRACE} \in \mathcal O^{k+1}$,
%\begin{align*}
%\left| \prob{k}{s_1}{\mathrm{TRACE}} - \prob{k}{s_2}{\mathrm{TRACE}} \right| \leq
%\varepsilon (k).
%\end{align*}
%\end{definition}
%}
%
\begin{definition}[$f(k)$-trace equivalence]
%\begin{definition}[$\varepsilon(k)$-trace equivalence]
Let ${\mchain = \mtuple}$ be a LMC. 
For a non-decreasing function $f: \mathbb N \rightarrow [0,1]$, 
we say that states $s_1,s_2 \in \msspace$ are $f(k)$-approximate probabilistic trace equivalent if for all $k \in \mathbb N$,
\begin{align*}
d_\mathrm{TV}\big(\prob{k}{s_1}{\cdot},\prob{k}{s_2}{\cdot}\big) \leq f(k),
\end{align*}
or alternatively if over 
$\mathrm{TRACE} \subseteq \mathcal O^{k+1}$,
\begin{align*}
\left| \prob{k}{s_1}{\mathrm{TRACE}} - \prob{k}{s_2}{\mathrm{TRACE}} \right| \leq f (k).
\end{align*}
%\qed
\end{definition}
%
%\alex{[I've reshuffled text and merged the remark within the paragraph. ]}
The condition on monotonicity follows from the requirement on the total variation distance, 
which is defined over a product output space and necessarily accumulates over time.   
The notion of $f(k)$-trace equivalence can be used to relate states from two different LMCs, 
much in the same way as $\varepsilon$-bisimulation. 

%Trivially, any two states are always related by $f$ being the Heaviside function, which is constant and equal to one for any $k \geq 0$.  
%We are therefore interested in seeking tight bounds on the approximation error in time, by synthesising a sensible function of time $f(k)$.  
One can introduce the notion of \emph{probabilistic trace distance} between pairs of states $s_1,s_2$ as
$$
\min\{f(k) \geq 0 \mid s_1 \text{ is } f(k)\text{-trace equivalent to }s_2\} 
= 
d_\mathrm{TV}\big(\prob{k}{s_1}{\cdot},\prob{k}{s_2}{\cdot}\big).  
$$ 
Notice that the RHS is clearly a pseudometric. 
We discuss the development of tight bounds on the probabilistic trace distance in Section \ref{sect: approx bisim implies approx trace equiv}. 

%\begin{remark}
%One can introduce the notion of probabilistic trace distance between states $s_1,s_2$ as
%$
%\min\{f(k) \geq 0 \mid s_1 \text{ is } f(k)\text{-trace equivalent to }s_2\} 
%= 
%d_\mathrm{TV}\big(\prob{k}{s_1}{\cdot},\prob{k}{s_2}{\cdot}\big) 
%$. 
%We discuss the development of tight bounds on the probabilistic trace distance in Section \ref{sect: approx bisim implies approx trace equiv}. 
%%\qed
%\end{remark} 

%The reason for the error tolerance of $1-(1-\varepsilon)^k$ is because they are the tightest bounds on the total variation of length $k$ traces guaranteed by $\varepsilon$-bisimulation, as we will see in Section \ref{sect: approx bisim implies approx trace equiv}. The error tolerance $1-(1-\varepsilon)^k$ is strictly increasing and allows the two processes to gradually accumulate total variation\footnote{Note that total variation of the trace is necessarily accumulated as a result of its increasing length, even though the total variation at the $k^\mathrm{th}$ step could abruptly go to $0$ if the processes have the same stationary distribution (see e.g. the phenomenon of ``cut-off'' \cite{AD86})}; for small values this increases like $k\varepsilon$, but it is bounded above by 1 (as well as by $k\varepsilon$).

\subsection{Interpretation and Application of $\varepsilon$-Trace Equivalence}
\label{sect:Interpretation of APTE}

The notion of $\varepsilon$-trace equivalence subsumes closeness of finite-time traces, 
and can be interpreted in two different ways. 
Firstly, 
$\varepsilon$-trace equivalence leads to closeness of satisfaction probabilities over bounded-horizon linear time properties,
%e.g. bounded reachability.
e.g. bounded LTL formulae, as follows.   
\begin{theorem}
Let $\mchain = \mtuple$ be an LMC,
and let $s_1,s_2 \in \msspace$ be $\varepsilon$-trace equivalent.
Let $\psi$ be any bounded LTL property over a $k$-step time horizon,
defined within the LTL fragment
$
\phi = \mathsf T \mid a \mid \phi \wedge \phi 
%\mid \phi_1 \vee \phi_2 
\mid \neg \phi \mid \phi \mathrm U^{\leq t} \phi 
$
for $t \leq k$.
Then,
\begin{align*}
\Bigm| \mathbb P \left[ s_1 \models \psi \right] - \mathbb P \left[ s_2 \models \psi \right] \Bigm|\; \leq f (k), 
%1-(1-\varepsilon)^k, 
\end{align*}
where $P \left[ s \models \psi \right]$ is the probability that starting from state $s$, the LMC satisfies property $\psi$. 
\begin{proof}
Formula $\psi$ is satisfied by a specific set of length $k+1$ traces. 
\qed 
\end{proof}
\end{theorem}
%
%More formally, each specification defined within the LTL fragment 
%$$
%\phi = \phi_1 \mathrm U^{\leq t} \phi_2 \mid a \mid \phi_1 \wedge \phi_2 \mid \phi_1 \vee \phi_2 %\mid \neg \phi_1, 
%$$ 
%for $t \leq k$, is satisfied by a specific set of length $k+1$ traces.   
%Given an LMC, let $\phi$ be any bounded 
%LTL property over a $k$-step time horizon. 
%Let states $s_1,s_2$ be $\varepsilon$-trace equivalent. Then,
%\begin{align*}
%\Bigm| \mathbb P \left[ s_1 \models \phi \right] - \mathbb P \left[ s_2 \models \phi \right] \Bigm|\; \leq f (k), 
%1-(1-\varepsilon)^k, 
%\end{align*}
%where $P \left[ s \models \phi \right]$ is the probability that starting from state $s$, the LMC satisfies property $\phi$. 
%
Alternatively, 
via its connection to the notion of total variation, 
$\varepsilon$-trace \emph{distance} leads to the notion of distinguishability of the underlying LMC, 
namely the ability (of an agent external to the model) to distinguish a model by observing its traces. 

%\alex{IMPORTANT: it seems like the statement above holds if you substitute $f(k)$ by the total variation distance. 
%In fact, you could always upper-bund the TV distance by a $f(k) = 1$ (Heavyside function), which would lead to a correct guess with probability 1. 
%The amended result is indeed what you have in the proof of the statement in the appendix. }
%\colin{
%[Thanks for spotting that error - should now be fixed by replacing ``trace equivalence'' with ``trace distance'']
%}

\begin{theorem}
\label{prop:TV and distinguishability} 
%\alex{[re-state wrt epsilon equivalence?]}
Let $s_1, s_2$ be two states of an LMC. 
Suppose one of them is selected by a secret fair coin toss. 
An external agent guesses which one has been selected by observing a trace of length $k+1$ emitted from the unknown state. 
Then, an optimal agent guesses correctly with probability 
\begin{align*}
%\frac{1}{2} + \frac{1}{2} d_\mathrm{TV}\big(\prob{k}{s_1}{\cdot},\prob{k}{s_2}{\cdot}\big)
\frac{1}{2} + \frac{1}{2} f(k), 
\end{align*}
with $f(k) = d_\mathrm{TV}\big(\prob{k}{s_1}{\cdot},\prob{k}{s_2}{\cdot}\big)$ being the probabilistic trace distance.  
%In particular, if $f(k)=0$ then the agent does no better than a random guess, since the two states are indistinguishable based on the traces they emit. 
%On the other hand, 
%if $f(k)=1$ then the agent will always be able to distinguish the two states, 
%with increased likelihood as time step $k$ grows. 
%\colin{since there is no set of traces that both states emit with non-zero probability.}
%prop
\end{theorem}
%\begin{proof}
%See Appendix \ref{Appendix: TV and distinguishability}.
%\end{proof}
%
%\begin{corollary}
%Suppose $s_1, s_2$ are $\varepsilon$-trace equivalent \alex{[$\varepsilon$-bisimilar]}. 
%Then an optimal agent will guess correctly with probability of at most
%$
%1 - \frac{(1-\varepsilon)^k}{2}. 
%$
%\end{corollary}
\section{$\varepsilon$-Probabilistic Bisimulation induces Approximate Probabilistic Trace Equivalence}
\label{sect: approx bisim implies approx trace equiv}

In this section we present the main result: 
we show that $\varepsilon$-bisimulation induces a tight upper bound on the probabilistic trace distance, 
quantifiable as $(1-(1-\varepsilon)^k)$. 
%namely that $\varepsilon$-bisimulation implies approximate probabilistic trace equivalence, for a specific choice of a function that depends on the parameter $\varepsilon$.   
This translates to a guarantee on all the properties implied by $\varepsilon$-trace equivalence, 
such as closeness of satisfaction probabilities for bounded linear time properties. 
In addition, since for finite state LMPs the maximal $\varepsilon$-bisimulation relation can be computed in $\mathcal O(\left| \msspace \right|^7)$ time \cite{DLT08}, 
this result allows to establish an upper bound on the probabilistic trace distance 
%in $\mathcal O(\left| \msspace \right|^7)$ 
with the same time complexity. 

\begin{figure}
\centering
%trim=left bottom right top
\includegraphics[clip, width=80mm, trim=5.5cm 18cm 4.75cm 3.35cm]{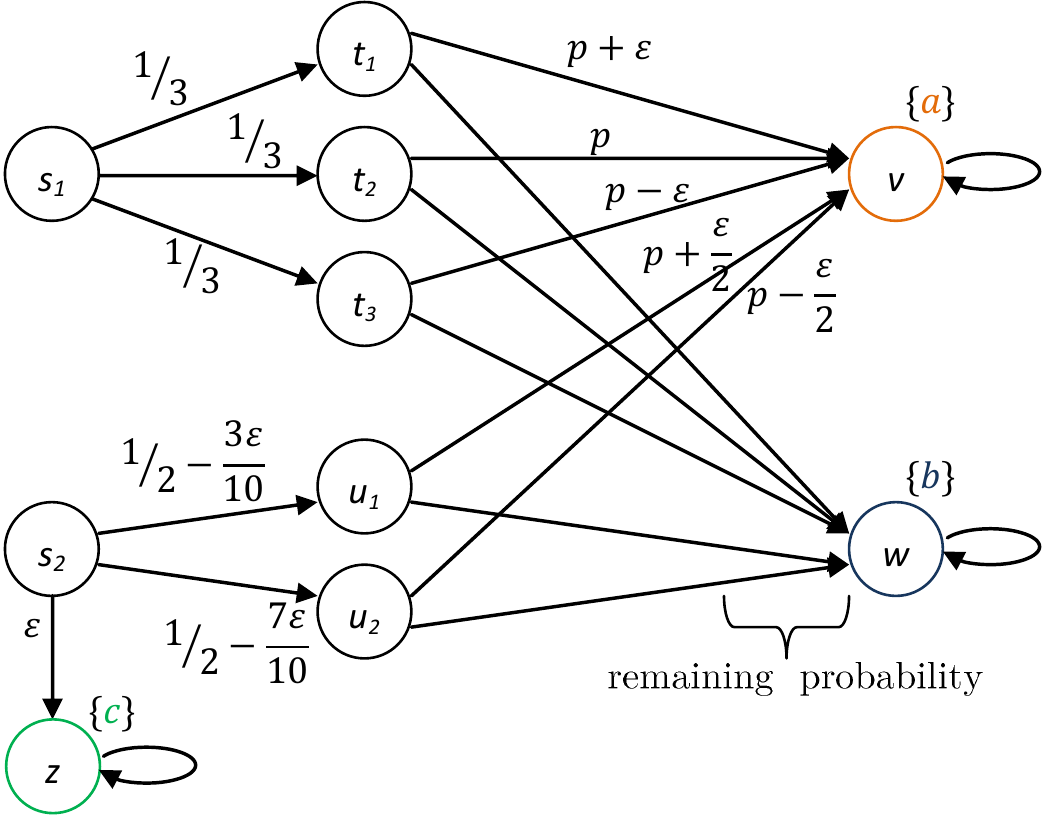} 
\caption{\textbf{LMC for the proof of Theorem \ref{them:approx bisim implies approx trace equiv}.}} 
\label{fig:proof}
\end{figure}

\begin{theorem}[$\varepsilon$-bisimulation implies $(1-(1-\varepsilon)^k)$-trace equivalence]
\label{them:approx bisim implies approx trace equiv}
Let $\mchain = \mtuple$ be a LMC.
If $s_1, s_2 \in \msspace$ are $\varepsilon$-bisimilar, then $s_1, s_2$ are $(1-(1-\varepsilon)^k)$-trace equivalent.
\begin{proof}[Sketch]
The full proof, 
developed for LMCs over uncountable state spaces,
can be found in Appendix \ref{Appendix: APB implies APTE}. 
Here we offer a sketch of proof, employing the finite-state LMP in Figure \ref{fig:proof} as an illustrating example 
(where for simplicity we have omitted the labels for internal states, 
which can as well be labelled with $\emptyset \in \mathcal O$). 
%The , but we converge on integrals with simple functions, and so the full proof will use the same ideas.

The maximal (i.e. coarsest) $\varepsilon$-bisimulation relation $R_{\varepsilon}$ is obtained by pairs of states within the sets 
\begin{align*}
%R_{\varepsilon} \coloneqq 
\{t_1, t_2, u_1 \},
%\cup 
\{t_2, t_3, u_2 \},
%\cup 
\{s_1, s_2 \}, 
%\alex{\{z\}, \{v, w\}},
\{v\},\{w\},\{z\}. 
\end{align*} 
%\alex{where we have disregarded to simplicity the labels over the internal states.}
We would like to prove that these $\varepsilon$-bisimilar states are also $\varepsilon$-trace equivalent. 
In the full proof, we will show this by induction on the length of the trace, for all $\varepsilon$-bisimilar states at the same time. 
In this sketch proof, we aim to illustrate the induction step by showing how to bound
\begin{align*}
\left| \prob{3}{s_1}{\lozenge^{\leq 3} a} - \prob{3}{s_2}{\lozenge^{\leq 3} a} \right|, 
\end{align*}
where $\lozenge^{\leq k} a$ is the set of traces of length $k+1$, 
which reach a state labelled with $a$ (which in this case is just state $v$).  
The idea is to match each of the outgoing transitions from $s_1$ to an outgoing transition from $s_2$ and to an $\varepsilon$-bisimilar state. 
Specifically, we explicitly write
\begin{equation}
\label{eqn:break up and match transitions1}
\begin{split}
&\prob{3}{s_1}{\lozenge^{\leq 3} a}
= \frac{1}{3} \prob{2}{t_1}{\lozenge^{\leq 2} a}
+ \frac{1}{6} \prob{2}{t_2}{\lozenge^{\leq 2} a}
\\&\indent\indent
+ \frac{1}{6} \prob{2}{t_2}{\lozenge^{\leq 2} a}
+ \frac{1}{3} \prob{2}{t_3}{\lozenge^{\leq 2} a}, 
\end{split}
\end{equation}
and respectively 
\begin{equation}
\begin{split}
\label{eqn:break up and match transitions2}
&\prob{3}{s_2}{\lozenge^{\leq 3} a}
= (\frac{1}{3}-\frac{3\varepsilon}{10}) \cdotp \prob{2}{u_1}{\lozenge^{\leq 2} a}
+ \frac{1}{6} \prob{2}{u_1}{\lozenge^{\leq 2} a}
\\&\indent\indent 
+ \frac{1}{6} \prob{2}{u_2}{\lozenge^{\leq 2} a}
+ (\frac{1}{3}-\frac{7\varepsilon}{10}) \cdotp \prob{2}{u_2}{\lozenge^{\leq 2} a}. 
\end{split}
\end{equation}
We then match-off the terms in the expansions for $\prob{3}{s_1}{\lozenge^{\leq 3} a}$ and $\prob{3}{s_2}{\lozenge^{\leq 3} a}$, one term at a time. We use the induction hypothesis to argue that the probabilities in the matched terms are $(1-(1-\varepsilon)^k)$-close to each other (here $k=1$), since they concern $\varepsilon$-bisimilar states. That is,
\begin{align*}
  & \left| \prob{2}{t_1}{\lozenge^{\leq 2} a} - \prob{2}{u_1}{\lozenge^{\leq 2} a} \right|
      \leq 1-(1-\varepsilon)^k
\\& \left| \prob{2}{t_2}{\lozenge^{\leq 2} a} - \prob{2}{u_1}{\lozenge^{\leq 2} a} \right|
      \leq 1-(1-\varepsilon)^k  
\\& \cdots
\end{align*}
where again $k = 1$. 
The total amount of difference between the matching coefficients is no more than $\varepsilon$. It can be shown (Lemma \ref{lemma:bound algebra} in Appendix \ref{Appendix: APB implies APTE}) that these conditions guarantee the required bound on $\left| \prob{3}{s_1}{\lozenge^{\leq 3} a} - \prob{3}{s_2}{\lozenge^{\leq 3} a} \right|$.

The main difficulty is choosing a suitable decomposition of $\prob{3}{s_1}{\lozenge^{\leq 3} a}$ and $\prob{3}{s_2}{\lozenge^{\leq 3} a}$.
%In particular, the $\nicefrac{1}{3}$ probability of transitioning into $t_2$ had to be broken up into two terms with $\nicefrac{1}{6}$ probability each in (\ref{eqn:break up and match transitions1}). A suitable decomposition can be generated using an extension of Hall's Matching Theorem \cite{BV75} (c.f. Appendix \ref{Appendix: APB implies APTE}).\qed 
This is non-trivial since in (\ref{eqn:break up and match transitions1}), the $\nicefrac{1}{3}$ probability of transitioning into $t_2$ had to be broken up into two terms with $\nicefrac{1}{6}$ probability each. However, we can tackle this issue using an extension of Hall's Matching Theorem \cite{BV75} (cf. Appendix \ref{Appendix: APB implies APTE}). It is then relatively straight forward to adapt this proof to LMCs on uncountable state spaces by converging on integrals with simple functions. \qed 
\end{proof}
\end{theorem} 

We now show that the expression for the induced bound on probabilistic trace distance proved in Theorem \ref{them:approx bisim implies approx trace equiv}, 
namely $(1-(1-\varepsilon)^k)$, is tight, in the sense that for any $k$ and $\varepsilon$, 
the bound can be attained by some pair of $\varepsilon$-bisimilar states in some LMC. 
In other words, 
it is not possible to provide bounds on the induced approximation level for traces, 
that are smaller than the expression discussed above and that are valid in general. 

%prop
\begin{theorem}
\label{prop:APTE gives the tightest bounds}
For any $\varepsilon \geq 0$, there exists a LMC $\mchain = \mtuple$ and states $s_1, s_2 \in \msspace$ such that $s_1, s_2$ are $\varepsilon$-bisimilar, and for all $k \in \mathbb N$ there exists a set $\mathrm{TRACE}$ of length $k+1$ traces s.t.
$\left| \prob{k}{s_1}{\mathrm{TRACE}} - \prob{k}{s_2}{\mathrm{TRACE}} \right| = 1-(1-\varepsilon)^k$.
\begin{proof}
Select $\varepsilon \geq 0$ and consider the following LMC: 
\begin{figure}[H]
\centering
%trim=left bottom right top
\includegraphics[clip, width=60mm, trim=3cm 23.5cm 10cm 4.5cm]{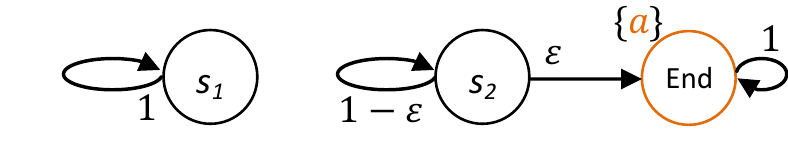}
\end{figure}
\noindent
Here $s_1, s_2$ are $\varepsilon$-bisimilar, 
and for all $k \in \mathbb N$, 
$\prob{k}{s_1}{\lozenge^{\leq k} a} = 0$, whereas 
$\prob{k}{s_2}{\lozenge^{\leq k} a} = \sum^k_{i=1} {\left(1-\varepsilon\right)^{i-1} \varepsilon} = 1-(1-\varepsilon)^k$.  
\qed 
\end{proof} 
\end{theorem}

The result in Theorem \ref{them:approx bisim implies approx trace equiv} can be viewed as an extension 
of the known fact that bisimulation implies trace equivalence in non-probabilistic transition systems. 
Similar to the deterministic case, the converse of Theorem \ref{them:approx bisim implies approx trace equiv} does not hold.

%prop
\begin{theorem}
\label{thm:noinverse}
$(1-(1-\varepsilon)^k)$-trace equivalence does not imply $\varepsilon$-bisimulation.
\begin{proof}
In Figure \ref{expl:branching time too strong}, states $s_1, s_2$ are not $\varepsilon$-bisimilar for any $\varepsilon < \nicefrac{1}{2}$, yet their probabilistic trace distance is equal to $0$. 
\qed
%In Example \ref{expl:branching time too strong}, states $s_1, s_2$ are not $\varepsilon$-bisimilar for any $\varepsilon < \nicefrac{1}{2}$ since none of the states $t, u_1, u_2$ are $\varepsilon$-bisimilar with each other. However, $s_1, s_2$ have a probabilistic trace distance equal to $0$.
\end{proof}
\end{theorem}

%This result further emphasises that although the $1-(1-\varepsilon)^k$ bound on probabilistic trace distance guaranteed by $\varepsilon$-bisimulation is tight as a uniform bound over the LMC (as discussed above), it is not tight for individual pairs of states. 
This example shows that $\varepsilon$-bisimulation cannot be used to effectively estimate the probabilistic trace distance between individual states. In particular, while the $(1-(1-\varepsilon)^k)$-bound on probabilistic trace distance discussed above is tight as a uniform bound, it is not tight for individual  pairs of states. 

\section{Application to Model Checking of Continuous-State LMCs}
\label{sect: application to model checking}

\newcommand{\mchainA}{\mchain^\mathcal A}
\newcommand{\msspaceA}{\msspace^\mathcal A}
\newcommand{\mmblA}{\mmbl^\mathcal A}
\newcommand{\mkerA}{\mker^\mathcal A}
\newcommand{\mlabelA}{\mlabel^\mathcal A}
\newcommand{\mtupleA}{{\left( \msspaceA, \mmblA, \mkerA, \mlabelA \right)}}
\newcommand{\sA}{s^\mathcal A}

\newcommand{\mchainC}{\mchain^\mathcal C}
\newcommand{\msspaceC}{\msspace^\mathcal C}
\newcommand{\mmblC}{\mmbl^\mathcal C}
\newcommand{\mkerC}{\mker^\mathcal C}
\newcommand{\mlabelC}{\mlabel^\mathcal C}
\newcommand{\mtupleC}{{\left( \msspaceC, \mmblC, \mkerC, \mlabelC \right)}}
\newcommand{\sC}{s^\mathcal C}

\newcommand{\mchainU}{\mchainC \oplus \mchainA}
\newcommand{\msspaceU}{\msspaceC \uplus \msspaceA}
\newcommand{\mmblU}{\mmbl^\oplus}
\newcommand{\mkerU}{\mker^\oplus}
\newcommand{\mlabelU}{\mlabel^\oplus}
\newcommand{\mtupleU}{\left( \msspaceU, \mmblU, \mkerU, \mlabelU \right)}

\newcommand{\partition}{\mathcal Q_\varepsilon}

Suppose we are given an LMC $\mchainC = \mtupleC$, 
which we shall refer to as the ``concrete'' model,  
possibly over a continuous state space. 
We are interested 
%and we want to calculate numerical results about its behaviour, e.g. 
in calculating its probability of satisfying a given LTL formula, starting from certain initial states.  
One approach is to construct a finite-state LMC $\mchainA = \mtupleA$ (the ``abstract model'') that can be related to $\mchainC$ (in a way to be made precise shortly).  
Probabilistic model checking can then be run over $\mchainA$ using standard tools for finite-state models such as PRISM \cite{KNP11}, 
and since $\mchainA$ is related to $\mchainC$, 
this leads to approximate outcomes that are valid for $\mchainC$. 
The above approach has been studied in several papers \cite{AKLP10, AKNP14, SA13}, and the method for constructing $\mchainA$ from $\mchainC$ is to raise smoothness assumptions on the kernel $\mker^\mathcal C$ of $\mchain^\mathcal C$, 
and to partition the state space $\msspace^\mathcal C$% sufficiently finely
, thus obtaining $\msspace^\mathcal A$ and $\mker^\mathcal A$ 
(the sigma algebra and labels being directly inherited).  
% so that $\mker^\mathcal C (\cdot, T)$ is almost constant inside each partition.

In this section we will demonstrate the application of our results.
We will employ $\varepsilon$-bisimulation to relate $\mchainA$ and $\mchainC$,
and use our results to bound their trace distance.
%We will employ $\varepsilon$-bisimulation to quantify the notion of $\mchainA$ having a ``similar structure'' to $\mchainC$, and the notion of probabilistic trace distance to quantify the approximation in the results. 
This method produces tighter error bounds, for a broader class of properties, than are currently established in literature.
The first step is to establish simpler conditions that guarantee $\varepsilon$-bisimulation between $\mchainC$ and $\mchainA$.

%prop
\begin{theorem}
\label{prop:partition for approx bisim}
Let $\varepsilon \in \left[0,1\right]$, and suppose there exists a finite measurable partition $\partition = \{ P_1, \cdots, P_N \}$ of $\msspaceC$ such that for all $P \in \partition$, $s_1, s_2 \in P$, we have that
$\mlabelC(s_1) = \mlabelC(s_2)$ 
and\footnote{
The left hand side is just $d_\mathrm{TV}(\mu_1,\mu_2)$ where for $i=1,2$, for $A \subseteq \partition$, $\mu_i(A) \coloneqq \mkerC \left( s_i, \bigcup A \right)$
}
\begin{align*}
\max_{J \subseteq \{1, \cdots, N\}} \left| \mkerC \left( s_1, \bigcup_{j \in J} P_j \right) - \mkerC \left( s_2, \bigcup_{j \in J} P_j \right) \right|
\leq \varepsilon. 
\end{align*}
\\\noindent
Assume wlog $P_i \neq \emptyset$, 
and for each $i \in \{ 1, \cdots, N \}$ choose a representative point $\sC_i \in P_i$. 
Consider the abstract model to be $\mchainA = \mtupleA$ formed by merging each $P_i$ into $\sC_i$. Formally,
\begin{itemize}
\item$\msspaceA = \{ \sA_1, \cdots, \sA_N \}$. 
\item$\mmblA = \mathcal P \left( \msspaceA \right)$. 
\item$\mkerA$ is such that $\mkerA(\sA_i, \{ \sA_j \}) = \mkerC( \sC_i, P_j )$. 
\item$\mlabelA(\sA_i) = \mlabelC(\sC_i)$. 
\end{itemize}
Then, for all $i \in \{ 1, \cdots, N \}$, $\sC \in P_i$, we have that $\sC$ is $\varepsilon$-bisimilar to $\sA_i$, 
and hence $1-(1-\varepsilon)^k$-trace equivalent. 
\end{theorem}

In practical terms the partition $\partition$ can be straightforwardly constructed in many cases. 
As shown in Theorem \ref{prop:smooth kernel}, the approach in \cite{AKLP10, AKNP14, IA13} 
%(which we restate below) 
generates a partition of $\msspaceC$ satisfying the conditions of Theorem \ref{prop:partition for approx bisim}. 
Thus, $\varepsilon$-bisimulation can be seen as the underlying reason for the closeness of 
%bounded reachability 
probabilities of events.

%prop
\begin{theorem}
\label{prop:smooth kernel}
Consider an LMC $\mchainC = \mtupleC$ where $\msspaceC$ is a Borel subset of $\mathbb R^d$.
Suppose that $\mkerC(s,T)$ is of the form $\int_{t \in T} f(s,t) \mathrm{d}t$, so that for each state $s \in \msspaceC$, $f(s,\cdot) : \msspaceC \to \mathbb R^+_0$ is the probability density of the next state. 
Suppose further that $f(\cdot,t)$ is uniformly $K$-Lipschitz continuous for all $t \in \msspaceC$. That is, for some $K \in \mathbb R$, for all $s_1, s_2, t \in \msspaceC$,
\begin{align*}
\left| f(s_1,t) - f(s_2,t) \right| \leq K \cdot \Vert s_1 - s_2 \Vert. 
\end{align*}
For $A \in \mmblC$ (so $A \subseteq \mathbb R^d$), let $\lambda(A)$ be the volume of $A$ and $\delta(A) \coloneqq \sup_{x_1, x_2 \in A} \{ \Vert x_1 - x_2 \Vert \}$ be the diameter of $A$.
For any $\varepsilon \in \left[0,1\right]$, finite $\lambda(\msspaceC)$, suppose partition $\mathcal Q = \{P_1, \cdots, P_N\}$ of $\msspaceC$ is such that
\begin{align*}
%\colin{
%\max_{j \in \{1, \cdots, N \}} \delta(P_j) \leq \frac{\varepsilon} {K\lambda(\msspaceC)}.
\max_{j \in \{1, \cdots, N \}} \delta(P_j) \leq \frac{2 \varepsilon} {K\lambda(\msspaceC)}.
%}
\end{align*}
Then, we have that $\mathcal Q$ satisfies the conditions of Theorem \ref{prop:partition for approx bisim} and can be used to construct the abstract model $\mchainA$.
\end{theorem}

There are a number of adaptations that could be made to this result.  
%, e.g. by only requiring the Lipschitz condition on $f(\cdot,t)$ to hold in each partition.
\cite{SA13} improves a related approach by varying the size of each partition in response to the local Lipschitz constant, rather than enforcing a globally uniform $K$. 
Similarly to this paper, \cite{AKNP14} also discusses the relation of approximate probabilistic bisimulation to the problem of generating the abstract model, but a strictly weaker definition of approximate probabilistic bisimulation is employed (cf. Section \ref{sect: alternative notions}).  
%which as we will show in Appendix \ref{sect: alternative notions}, does not in fact guarantee closeness of reachability probabilities (contrary to what is stated in \cite{AKNP14}).  
Finally, note that using algorithms in \cite{DLT08}, we can compute $\varepsilon$-bisimulation relations on $\mchainA$: 
this allows $\mchainA$ to be further compressed (at the cost of an additional $\varepsilon_2$ approximation), 
by merging the states that are $\varepsilon_2$-bisimilar to each other. 
% (like what we did in Theorem \ref{prop:partition for approx bisim}).
%However, since current model checking techniques are polynomial in the size of the state space for LTL properties, we will not further explore this idea.

\section{Case Study}
\label{sect:application case study}

\textbf{Concrete Model. }
Consider the concrete model $\mchainC = \mtupleC$, describing the weather forecast for a 
%remote holiday 
resort. Here $\msspaceC = \{0,1\} \times \left[0,1\right)$, $\mmblC = \mathcal B(\msspaceC)$, 
and the state at time $t$ is $(R_t, H_t)$, where
\begin{itemize}
\item $R_t \in \{0,1\}$ is a random variable representing whether it rains on day $t$, 
\item $H_t \in \left[0,1\right)$ is a random variable representing the humidity after day $t$. 
\end{itemize}

Raining on day $t$ causes it to become more likely to rain on day $t+1$, but it also tends to reduce the humidity, which causes it to become gradually less likely to rain in the future. 
%as the humidity gets used up. 
The meteorological variations are encompassed by $\mkerC$, 
which is such that the model evolves according to 
\begin{equation*}
\begin{split}
&\mathbb P ( R_{t+1} \mid R_0, \cdots, R_t, H_0, \cdots, H_t ) =  \mathbb P ( R_{t+1} \mid R_t, H_t )
\\&\indent\indent \sim \begin{cases}
  \mathrm B(\frac{1}{4} + \frac{3}{4} H_t) & \text{if}\ R_t=1 \\
  \mathrm B(\frac{3}{4} H_t) & \text{if}\ R_t=0
\end{cases},
\\
&\mathbb P( H_{t+1} \mid R_0, \cdots, R_t, H_0, \cdots, H_t, R_{t+1} ) = \mathbb P( H_{t+1} \mid H_t, R_{t+1} )
\\&\indent\indent \sim \begin{cases}
  \mathrm U \left[0, \frac{1+H_t}{2} \right) & \text{if}\ R_{t+1}=1 \\
  \mathrm U \left[\frac{H_t}{2}, 1 \right) & \text{if}\ R_{t+1}=0
\end{cases}, 
\end{split}
\end{equation*}
where $\mathrm B(p)$ is the Bernoulli distribution with probability $p$ of producing $1$, and $\mathrm U \left[a,b \right)$ is the uniform distribution over the real interval $\left[a,b \right)$. 
Finally, the states of the model are labelled according to whether it rains on that day, namely 
\begin{align*}
\mlabelC((r,h)) = \begin{cases}
  \{\mathrm{RAIN}\} & \text{if}\ r = 1 \\
  \emptyset & \text{if}\ r = 0. 
\end{cases}
\end{align*}
Given $\mchainC$ we are interested in computing the likelihood of events expressing meteorological predictions, 
given knowledge of present weather conditions.   

\medskip

\noindent\textbf{Synthesis of the Abstract Model.} 
%
%In this case, $\mchainC$ does not satisfy the smoothness assumptions of Theorem \ref{prop:smooth kernel}: firstly, $\mkerC$ does not admit a density because of the discrete coordinate; secondly, even if we condition on $R_{t+1} = 1$, the conditional density $f_{H_{t+1} \mid H_t, R_t, R_{t+1}=1} (h_{t+1}) = \frac{2}{1+H_t} \mathds{1}_{\left[0,\frac{1+H_t}{2}\right)} (h_{t+1})$ is discontinuous in $h_{t+1}$. Nonetheless, we can still construct $\mchainA$ by taking a sensible partition of $\msspaceC$ and proving that it satisfies the conditions of Theorem \ref{prop:partition for approx bisim}. 
Notice that $\mchainC$ does not directly satisfy the smoothness assumptions of Theorem \ref{prop:smooth kernel}, 
in view of the discrete/continuous structure of its state space and the discontinuous probability density resulting from the uniform distribution. 
%associated stochastic kernel. 
Nonetheless, we can still construct $\mchainA$ by taking a sensible partition of $\msspaceC$ and proving that it satisfies the conditions of Theorem \ref{prop:partition for approx bisim}. 
Let $\partition \coloneqq \big\{ P_{r,h} \bigm| r \in \{0,1\}, \; h \in \{0,\cdots,N-1\} \big\}$, where $ P_{r,h} = \{r\} \times \left[\nicefrac{h}{N},\nicefrac{h+1}{N}\right) $.

%prop
\begin{theorem}
\label{thm:casestudy}
For any $\varepsilon \in \left[0,1\right]$, by taking $N \geq \nicefrac{2}{\varepsilon}$, we have that $\partition$ satisfies the conditions of Theorem \ref{prop:partition for approx bisim}.
\end{theorem}

%\begin{proof}
%For any $r\in\{0,1\}$, $h\in\{0,\cdots,N-1\}$ and $s_1,s_2 \in P_{r,h}$, assume wlog $s_1 = (r, x_1), s_2 = (r,x_2)$ and $x_1 \leq x_2$. Then, $\mkerC (s_1, \cdot), \mkerC (s_2, \cdot)$ differ only on $V \subseteq \msspaceC$ where
%\begin{align*}
%V = \begin{cases}
%  \{0,1\} \times \left[ \frac{1+x_1}{2}, \frac{1+x_2}{2} \right) & \text{if}\ r=1 \\
%  \{0,1\} \times \left[ \frac{x_1}{2}, \frac{x_2}{2} \right) & \text{if}\ r=0
%\end{cases}
%\end{align*}
%Therefore, we have for all $T \in \mmblC$,
%\begin{align*}
%  &        \left| \mkerC (s_1, T) - \mkerC (s_2, T) \right|
%\\&\indent\indent \leq \sup_{s\in\msspaceC} \mkerC(s,V)
%\\&\indent\indent \leq \sup_{p_\mathrm{rain} \in \left[0,1\right]} \Big| p_\mathrm{rain} (x_2-x_1)+ (1-p_\mathrm{rain}) (x_2-x_1) \Big|
%\\&\indent\indent = \nicefrac{1}{N}
%\end{align*}
%\end{proof}
%prop

Therefore, we may construct the abstract model using Theorem \ref{prop:partition for approx bisim}.
We choose the abstract state $(r^\mathcal A, h^\mathcal A) \in \msspaceA := \{0,1\} \times \{0, \cdots, N-1 \}$ to correspond to the partition $P_{r^\mathcal A, h^\mathcal A} \in \partition$, 
and within each partition we select the concrete state with the lowest $H_t$-coordinate to be the representative state. This produces the abstract model $\mchainA = \mtupleA$, 
where $\mmblA = \mathcal P(\msspaceA)$, $\mlabelA((r,h)) = \mlabelC((r,h))$, and 
%\begin{itemize}
%\item $
%\msspaceA = \{0,1\} \times \{0, \cdots, N-1 \}
%$
%\item $
%\mmblA = \mathcal P(\msspaceA)
%$
%\item 
\begin{align*}
\mkerA \left( (h_0,r_0), \{(h_1,r_1)\} \right)
 = \begin{cases}
  p_{\mathrm{R}_1 \mid \mathrm{R_0}}(h_0) \cdotp p_{H_1 \mid \mathrm{R}_1}(h_0,h_1)
& \text{if}\ r_0=1, r_1=1 \\
  \left( 1 - p_{\mathrm{R}_1 \mid \mathrm{R_0}}(h_0) \right) \cdotp p_{H_1 \mid \neg \mathrm{R}_1}(h_0,h_1)
& \text{if}\ r_0=1, r_1=0 \\
  p_{\mathrm{R}_1 \mid \neg\mathrm{R_0}}(h_0) \cdotp p_{H_1 \mid \mathrm{R}_1}(h_0,h_1)
& \text{if}\ r_0=0, r_1=1 \\ 
  \left( 1 - p_{\mathrm{R}_1 \mid \neg\mathrm{R_0}}(h_0) \right) \cdotp p_{H_1 \mid \neg \mathrm{R}_1}(h_0,h_1)
& \text{if}\ r_0=0, r_1=0, \\
\end{cases}
\end{align*}
%\item $
%\mlabelA((r,h)) = \begin{cases}
%  \{\mathrm{RAIN}\} & \text{if}\ r = 1 \\
%  \emptyset & \text{if}\ r = 0
%\end{cases}
%$
%\end{itemize}
where
\begin{itemize}
\item 
$p_{\mathrm{R}_1 \mid \mathrm{R_0}}(h_0) = \frac{1}{4} + \frac{3}{4} \frac{h_0}{N}$, 
and
%\item
$p_{\mathrm{R}_1 \mid \neg \mathrm{R_0}}(h_0) = \frac{3}{4} \frac{h_0}{N}$, 
\item
$p_{H\mid\neg\mathrm{R}} (h_0,h_1)
 = \frac{2}{2N-h_0} \cdotp \max \left( \min \left( h_1 + 1 - \nicefrac{h_0}{2}, 1 \right) , 0 \right)$,
\item
$p_{H\mid\mathrm{R}} (h_0,h_1) = \frac{2}{N+h_0} \cdotp \max \left( \min \left( \frac{N+h_0}{2} - h_1, 1 \right) , 0 \right)$.
\end{itemize}

\medskip

\noindent\textbf{Computation of Approximate Satisfaction Probabilities. }
Suppose that at the end of day 0, we have $R_0 = 0, H_0 = 0.5$, and a travel agent wants to know the risk of there being 2 consecutive days of rain over the next three days.

%\alex{[I've eliminated the pieces of PRISM code from the text.]}
%\colin{[Moved to appendix instead]}

This probability can be computed algorithmically according to $\mchainA$,
and for $N=1000$, this is $0.365437$ (see Appendix \ref{Appendix: PRISM code}).
Since point $(0, 500) \in \msspaceA$ is $0.001$-bisimilar with $(0,0.5) \in \msspaceC$, 
this means that according $\mchainC$ with initial state $(0,0.5) \in \msspaceC$, the probability of there being two consecutive days of rain over the next three days is
$0.365437 \pm 0.003$.
\\\\
\noindent\textbf{Analytical Validation of the Result.} 
In this setup it is possible to evaluate the exact result for $\mchainC$ analytically: 
\begin{align*}
\mathbb P \left[ R_1 = 1, R_2 = 1 \right]
 & = \mathbb P \left[ R_1 = 1 \right] \int_0^1 f_{H_1 \mid R_1 = 1}(h_1)
 \cdotp \mathbb P \left[ R_2 = 1 \mid H_1=h_1, R_1=1 \right] \mathrm{d}h_1, 
%\\& = 0.199219, 
\end{align*}
which amounts to $0.199219$, 
and similarly
%\begin{align*}
$\mathbb P \left[ R_1 = 0, R_2 = 1, R_3 = 1 \right] = 0.166626$. 
%\end{align*}
This yields 
\begin{align*}
& \mathbb P \left[ \text{Two consecutive rainy days over next 3 days} \right] = \\
& \indent \mathbb P \left[ R_1 = 1, R_2 = 1 z\right]  + \mathbb P \left[ R_1 = 0, R_2 = 1, R_3 = 1 \right]
= 0.365845,  
\end{align*}
which is within the error bounds guaranteed by $\varepsilon$-trace equivalence, as expected.

% filter(state, P=? [ (X (r=0)) & (X X (r=1)) & (X X X (r=1)) ], r=0&h=500)
% Me: \frac{1}{1024} \left( -9H_0^3 - 69 H_0^2 - 27 H_0 + 27 \right)
% Me at x = 0.25 -> 0.193405    (PRISM: 0.19308)
% Me at x = 0.5  -> 0.166626    (PRISM: 0.16635)
% Me at x = 0.75 -> 0.129776    (PRISM: 0.12958)

% filter(state, P=? [ (X (r=1)) & (X X (r=1)) ], r=0&h=500)
% Me: 1/64 * (21 H0 + 9 H0^2)
% Me at 0.25 -> 0.0908203    (PRISM: 0.09075)
% Me at 0.5  -> 0.199219     (PRISM: 0.19907)
% Me at 0.75 -> 0.325195     (PRISM: 0.32498)

\section{Other Notions of $\varepsilon$-Bisimulation}
\label{sect: alternative notions}

The following condition appears in literature \cite{A13, AKNP14, IAK12} as the definition of approximate probabilistic bisimulation. 
%It is strictly weaker than our notion of $\varepsilon$-bisimulation. 
% 
For simplicity, let us restrict our attention to finite state spaces.

%cond
\begin{definition}[Alternative notion of approximate probabilistic bisimulation, adapted from \cite{AKNP14}]
\label{cond:R-Closed Approx Bisim}
Let $\mchain = \mtuple$ be a LMC, where $\msspace$ is finite and $\mmbl = \mathcal P(\msspace)$.
For $\varepsilon \in [0,1]$, a binary relation $R_\varepsilon$ on $\msspace$ satisfies Definition \ref{cond:R-Closed Approx Bisim} if:
\begin{align*}
  & \forall (s_1, s_2) \in R_\varepsilon,
\indent\text{we have } \mlabel(s_1) = \mlabel(s_2),
\\& \forall (s_1, s_2) \in R_\varepsilon, \; \forall \tilde T \subseteq \msspace \text{ s.t. $\tilde T$ is $R_\varepsilon$-closed},
\indent\text{we have } \left| \mker(s_1, \tilde T) -  \mker(s_2, \tilde T)\right| \leq \varepsilon.
\end{align*}
%cond
\end{definition}

This is different from our notion of approximate probabilistic bisimulation because $\tilde T$ ranges over $R_\varepsilon$-closed sets rather than all (measurable) sets. This definition is closer to exact probabilistic bisimulation (cf. Definition \ref{def:Exact probabilistic bisimulation}), but it is too weak to effectively bound probabilistic trace distance.

\begin{theorem}
\label{prop:r-closed approx bisim does not imply approx trace equiv}
For any $\varepsilon > 0$, there exists an LMC $\mchain = \mtuple$, a binary relation $R_\varepsilon$ on $\msspace$, and a pair of states $(s_1,s_2) \in R_\varepsilon$, such that $R_\varepsilon$ satisfies the conditions in Definition \ref{cond:R-Closed Approx Bisim}, but the 2-step reachability probabilities from $s_1,s_2$ differ by $1$ for some destination states.
\begin{proof}
For $\varepsilon > 0$, let $N\in\mathbb{Z}^+$, $1/N \leq \varepsilon$. Consider the following LMC. Let $R_\varepsilon \coloneqq \{(s_1,s_2)\} \cup \{ (t_k,t_{k+1}) \mid \;k \in \{ 0, \cdots, N-1 \} \}$.
\begin{figure}[H]
\centering
%trim=left bottom right top
\includegraphics[clip, width=80mm, trim=4.5cm 18.25cm 5.75cm 5cm]{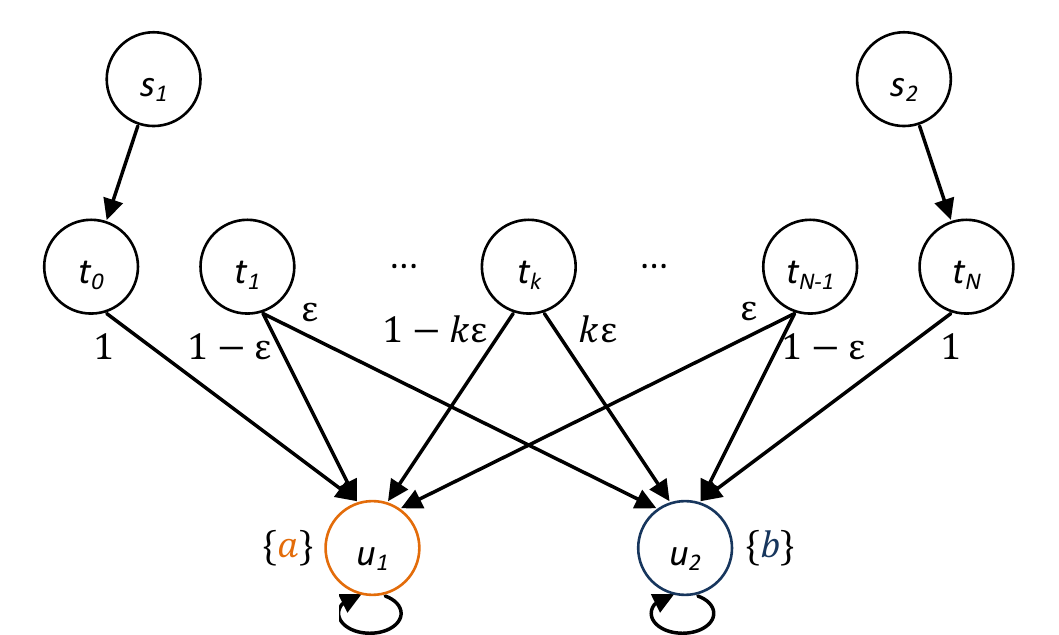}
\end{figure}
The only $R_\varepsilon$-closed sets are $\{s_1, s_2\}$, $\{t_0, \dots , t_N\}$, $\{u_1\}$, $\{u_2\}$ and unions of these sets, and so $R_\varepsilon$ satisfies Definition \ref{cond:R-Closed Approx Bisim}.

We have $s_1 R_\varepsilon \, s_2$, and yet
%\begin{align*}
$\prob{2}{s_1}{\lozenge^{\leq 2} a} = 1$
but $\prob{2}{s_2}{\lozenge^{\leq 2} a} = 0$, 
%\end{align*}
where $\lozenge^{\leq 2} a$ is the set of length $3$ traces that reach a state labelled with $a$.
\end{proof}
\end{theorem}

As shown in \cite{IAK12} however, there is still some relationship between the probabilities of specific traces, which hinges on additional details of the structure of the transition kernel. 

\section{Conclusions and Extensions}

In this paper we have developed a theory of $f(k)$-trace equivalence. We derived the minimum $f(k)$ such that $\varepsilon$-bisimulation implies $f(k)$-trace equivalence, thus extending the well known result for the exact non-probabilistic case.
By linking error bounds on the total variation of length $k$ traces to a notion of approximation based on the underlying transition kernel, 
we provided a means of computing upper bounds for the total variation and of synthesising abstract models with arbitrarily small total variation to a given concrete model.

It is of interest to extend our results to allow the states of the LMC to be labelled with bounded real-valued rewards, 
and then to limit the difference in expected reward between approximately bisimilar states.

%[Mention simulation vs bi-simulation here if we trim it from the main text].
%[Long shot:] As another possible extension, one could allow a metric $d$ over the state space $\msspace$, and use approximate bisimulation (or a variation of it) to bound the divergence between the traces between different states with respect to this metric. One possible way of quantifying this divergence is the expected distance between between processes starting at the two states under the minimum coupling. Our current theorem effectively covers the case where $d(s_1,s_2) = 0$ for $\mlabel(s_1)=\mlabel(s_2)$ and $d(s_1,s_2) = 1$ for $\mlabel(s_1) \neq \mlabel(s_2)$.

\section*{Acknowledgments} 
%\medskip
%\noindent\textbf{Acknowledgments.}
The authors would like to thank Marta Kwiatkowska for discussions on an earlier version of this draft. 

%% Either use bibtex (recommended), 
\bibliography{bibliography}

%% .. or use the thebibliography environment explicitely

%%% Appendix 

\newpage 

\appendix

\section{Proof of Theorem \ref{thm:0bisim}}
\label{Appendix: proof of 0-bisim}
\begin{proof}
Suppose $s_1,s_2$ are $0$-bisimilar. Let $R_\varepsilon$ be a $0$-bisimulation relation s.t. $s_1 R_\varepsilon s_2$,
and assume wlog that $R_\varepsilon$ is reflexive (by enlarging it if necessary).

Let $R^\infty$ be the transitive closure of $R_\varepsilon$, so that $R^\infty$ is an equivalence relation.
Since $s_1 R^\infty s_2$, we only need to show that $R^\infty$ is an exact probabilistic bisimulation relation.
The other conditions being easy to check, we will only prove $\mker(s_1, \tilde T) = \mker(s_2, \tilde T)$ for all $(s_1,s_2) \in R^\infty$ for all $R^\infty$-closed $\tilde T \in \mmbl$.

Let $(s_1,s_2) \in R^\infty$, and let $\tilde T \in \mmbl$ be $R^\infty$-closed. Then, $s_1 R_\varepsilon y_1 R_\varepsilon \cdots R_\varepsilon y_N R_\varepsilon s_2$ for some $y_1, \cdots, y_N \in \msspace$.
We have
$\mker(s_1, \tilde T) \leq \mker(y_1, R_\varepsilon(\tilde T))$ since $R_\varepsilon$ is a $0$-bisimulation relation,
and we have
$\mker(y_1, R_\varepsilon(\tilde T)) = \mker(y_1, \tilde T)$ since $R_\varepsilon(\tilde T) = \tilde T$.
We therefore have $\mker(s_1, \tilde T) = \mker(y_1, \tilde T)$ since $R_\varepsilon$ is symmetric. Similarly, we have $\mker(y_1, \tilde T) = \mker(y_2, \tilde T) = \cdots = \mker(y_N, \tilde T) = \mker(s_2, \tilde T)$. 
\qed
\end{proof}

\section{Proof of Theorem \ref{prop:TV and distinguishability}}
\label{Appendix: TV and distinguishability}
%\alex{
%Here you prove the result only for the TV distance. In the text below, I would eliminate the italicised paragraph (it is just a re-statement of the main proposition), and
%....I would recommend merging the two lemmas within the main proof, just having a straight proof of the statement. }
%\colin{PTAL}
\begin{proof} 
Let ${\mchain = \mtuple}$ be an LMC and $s_1,s_2 \in \msspace$. Let $\mchain$ start at the initial state $s_C$ where $C$ is a uniform random variable taking values in $\{1,2\}$. Conditional on $C = c$, the length $k+1$ trace emitted from $\mchain$ is a random variable $X$ taking values in $\mathcal O^{k+1}$ with distribution $\prob{k}{s_c}{\cdot}$.

After observing outcome $X=x$, an optimal agent guesses $C$ to be $1$ if $\mathbb{P}[C=1|X=x]>\mathbb{P}[C=2|X=x]$, and $C$ to be $2$ if $\mathbb{P}[C=2|X=x]>\mathbb{P}[C=1|X=x]$, in order to maximize the probability of being correct. So,
\begin{equation*}
\begin{split}
\mathbb{P}[&\mathrm{CORRECT}|X=x]
\\&= \max\{{\mathbb{P}[C=1|X=x]},{\mathbb{P}[C=2|X=x]}\}
\\&= \frac{1}{2} \Big( \mathbb{P}[C=1|X=x] + \mathbb{P}[C=2|X=x]
     + \big| \mathbb{P}[C=1|X=x] - \mathbb{P}[C=2|X=x] \big| \Big)
\\&= \frac{1}{2} + \frac{1}{2} \Big| \mathbb{P}[C=1|X=x] - \mathbb{P}[C=2|X=x] \Big|
\\&= \frac{1}{2} + \frac{
   \big| \mathbb{P}[X=x|C=1] - \mathbb{P}[X=x|C=2] \big|
}{
   4 \mathbb{P}[X=x]
}
\\&\indent\indent \text{since $\mathbb{P}[C=1|X=x] = \mathbb{P}[X=x|C=1]\frac{\mathbb{P}[C=1]}{\mathbb{P}[X=x]} = \frac{\mathbb{P}[X=x|C=1]}{2\mathbb{P}[X=x]}$}
\\&\indent\indent \text{and similarly for $\mathbb{P}[C=2|X=x]$.}
\end{split}
\end{equation*}
We obtain
%\allowdisplaybreaks
%\begin{flalign*}
\begin{equation*}
\begin{split}
\mathbb{P}[&\mathrm{CORRECT}]
\\&= \sum_{x \in \mathcal O^{k+1}} \mathbb{P}[\mathrm{CORRECT}|X=x] \mathbb{P}[X=x]
\\&= \frac{1}{2} + \frac{1}{4} \sum_{x \in \mathcal O^{k+1}} {\Big| \mathbb{P}[X=x|C=1]-\mathbb{P}[X=x|C=2]\Big|}
\\&= \frac{1}{2} + \frac{1}{2} \max_{\mathrm{TRACE} \subseteq \mathcal O^{k+1}}
       {\Big| \mathbb{P}[X \in \mathrm{TRACE} | C=1] - \mathbb{P}[X \in \mathrm{TRACE} | C=2]\Big|}
\\&= \frac{1}{2} + \frac{1}{2} d_\mathrm{TV}\big(\prob{k}{s_1}{\cdot},\prob{k}{s_2}{\cdot}\big). 
\end{split}
\end{equation*}
\qed
\end{proof}

\section{Proof of Theorem \ref{them:approx bisim implies approx trace equiv}}
\label{Appendix: APB implies APTE}

We say that $(Z, \mathcal F, \mu)$ is a measure space if $Z$ is a set, $\mathcal F \subseteq \mathcal P (Z)$ is a $\sigma$-algebra over $Z$ and $\mu : \mathcal F \to \mathbb R^+_0$ is a measure function. We say that $S \in \mathcal F$ is an atom if $\mu(S)>0$,  and if for any measurable subset $T \subseteq S$, we have either $\mu(T) = \mu(S)$ or $\mu(T) = 0$. For instance, for any measurable singleton point $\{x\} \in \mathcal F$, if $\mu(\{x\}) > 0$ then \{x\} is an atom. We borrow the following fact. 

\begin{theorem}[\cite{BV75}]
\label{them:Bollobas}
Let $(Z, \mathcal F, \mu)$ be an atomless measure space. Let $A$ be an index set, $(X_{\alpha})_{\alpha \in A}$ be a family of measurable sets in $\mathcal F$ of finite measure, and $(\lambda_\alpha)_{\alpha \in A}$ be a family of non-negative real numbers. The following holds:\\ 
There exists a family of measurable sets $(Y_\alpha)_{\alpha \in A}$ in $\mathcal F$ such that $\forall \alpha \in A$ we have $Y_\alpha \subseteq X_\alpha$, $\mu (Y_\alpha) = \lambda_\alpha$, and $\forall \alpha,\alpha' \in A$ with $\alpha \neq \alpha'$ we have $\mu(Y_\alpha \cap Y_{\alpha'}) = 0$
$$\iff$$
For every finite subset $B \subseteq A$,
%\begin{equation*}
%\label{eqn:Bollobas condition}
$\mu\left(\bigcup_{\alpha \in B} X_\alpha\right) \geq \sum_{\alpha \in B} \lambda_{\alpha}$.  
%\end{equation*}
\end{theorem}
We will also use the following lemma, where $u_i \cdot x_i$ corresponds to the terms in Equation (\ref{eqn:break up and match transitions1}) in the sketch proof (e.g. $u_1 = \nicefrac{1}{3}$, $x_1 = \prob{2}{t_1}{\lozenge^{\leq 2} a}$), and similarly $v_i \cdot y_i$ corresponds to the terms in Equation (\ref{eqn:break up and match transitions2}).
\begin{lemma}
\label{lemma:bound algebra}
Let $\varepsilon, \delta \geq 0$.
Let $u_i, v_i, x_i, y_i \in [0,1]$ for $i \in \{1, 2, \dots, M\}$.
Suppose that 
\begin{align}
  &\left|x_i - y_i\right| \leq \delta \;\; \forall i \in \{1, \dots, M\}, \label{eqn:lemma1}
\\&\sum^M_{i=1} u_i \leq 1, \label{eqn:lemma2}
\\&\sum_{i \mid u_i > v_i } \left( u_i-v_i \right) \leq \varepsilon. \label{eqn:lemma3}
\end{align}
Then
%\begin{align*}
$\sum^M_{i=1}{x_i u_i} - \sum^M_{i=1}{y_i v_i} \leq \varepsilon + \delta - \varepsilon\delta$. 
%\end{align*}
\begin{proof}
Let $y'_i \coloneqq \max\{0, x_i-\delta\}$, so that $y_i \geq y'_i$ by ($\ref{eqn:lemma1}$) and by $y_i \in \left[0,1\right]$.
Then,
\begin{align*}
\sum^M_{i=1}{x_i u_i} - \sum^M_{i=1}{y_i v_i}
& \leq \sum^M_{i=1}{x_i u_i} - \sum^M_{i=1}{y'_i v_i}
%\\& 
= \sum^M_{i=1}{u_i \left(x_i-y'_i\right)} + \sum^M_{i=1} {y'_i \left(u_i-v_i \right)}. 
\end{align*}
Now,
\begin{align*}
\sum^M_{i=1}{u_i \left(x_i-y'_i\right)}
& = \sum^M_{i=1} u_i \min\{x_i, \delta\}
%\\& 
\leq \delta \sum^M_{i=1} u_i
\\& \leq \delta \indent\indent\text{by ($\ref{eqn:lemma2}$)}. 
\end{align*}
We conclude that 
\begin{align*}
\sum^M_{i=1} {y'_i \left(u_i-v_i \right)}
& \leq \sum_{i \mid u_i > v_i \text{ and } y'_i \neq 0} {y'_i \left(u_i-v_i \right)}
\\& \leq \left( 1 - \delta \right) \sum_{i \mid u_i > v_i \text{ and } y'_i > 0} {\left(u_i-v_i\right)}
\\& \indent\indent \text{since for $i$ s.t. $y'_i \neq 0$, we have $y'_i = x_i - \delta \leq 1 - \delta$}
\\& \leq \left( 1 - \delta \right) \varepsilon \indent\indent\text{by ($\ref{eqn:lemma3}$)}. 
\end{align*}
\qed
\end{proof}
\end{lemma}

\noindent\textbf{Main Proof.} We are now ready for the proof of the main theorem. Let $\mchain = \mtuple$ be a LMC, let $\varepsilon \in \left[0,1\right]$, let $R_\varepsilon$ be an $\varepsilon$-bisimulation relation on $\mchain$. 
We prove by induction on $k \in \mathbb N$ that for any $(s_1, s_2) \in R_\varepsilon$, we have that 
\begin{align*}
d_{\mathrm{TV}}\left( \prob{k}{s_1}{\cdot}, \prob{k}{s_2}{\cdot} \right) \leq 1 - (1- \varepsilon)^k. 
\end{align*}
For the base case $k=0$, since $s_1 R_\varepsilon s_2$, we have $L(s_1)=L(s_2)$. Let $\alpha_0 \coloneqq L(s_1)=L(s_2)$, then 
\begin{align*}
\prob{0}{s_1}{\mathrm{TRACE}} = \prob{0}{s_2}{\mathrm{TRACE}}
= \begin{cases}
1 &\mbox{if } <\alpha_0>\in\mathrm{TRACE} \\ 
0 & \mbox{otherwise}
\end{cases} 
\end{align*}
For the induction step, we assume (as the induction hypothesis) that for a fixed $k-1 \in \mathbb N$, for any $(s_1, s_2) \in R_\varepsilon$,
\begin{align*}
d_{\mathrm{TV}}\left( \prob{k-1}{s_1}{\cdot}, \prob{k-1}{s_2}{\cdot} \right) \leq \delta, 
\end{align*}
where $\delta \coloneqq 1 - (1- \varepsilon)^{k-1}$. We will prove the result for $k$ by showing for any fixed $\mathrm{TRACE} \subseteq \mathcal O^{k+1}$,
\begin{align*}
\prob{k}{s_2}{\mathrm{TRACE}} \geq \prob{k}{s_1}{\mathrm{TRACE}} - \left(1-\left(1-\varepsilon\right)^k\right). 
\end{align*}
Since $R_\varepsilon$ is symmetric, we can reverse the roles of $s_1$ and $s_2$ to get the required result.

Let $\alpha \coloneqq \mlabel (s_1)$. Let $\mathrm{TRACE}_{\langle \alpha \rangle} \coloneqq \{ \langle \alpha_1, \cdots, \alpha_k \rangle \mid \langle \alpha, \alpha_1, \cdots, \alpha_k \rangle \in \mathrm{TRACE} \}$ be the set of traces from $\mathrm{TRACE}$ after seeing an initial $\alpha$. By conditioning on the first step, we have (from Definition \ref{def:LMC} and Definition \ref{def:ProbTrace}):
\begin{align*}
\prob{k}{s_1}{\mathrm{TRACE}} = \int_{y\in\msspace} {\prob{k-1}{y}{\mathrm{TRACE}_{\langle \alpha \rangle}} \mker(s_1,\mathrm{d}y)}. 
\end{align*}
Let $T(y) \coloneqq \prob{k-1}{y}{\mathrm{TRACE}_{\langle \alpha \rangle}}$. From the construction of the integral, this becomes
\begin{equation*}
\begin{split}
& \prob{k}{s_1}{\mathrm{TRACE}}
\\&\indent\indent = \sup \bigg\{ \sum^K_{r=1} c_r \mker(s_1, G_r) \;\bigg|\; \sum^K_{r=1} c_r \mathds{1}_{G_r}(t) \leq T(t) \text{ and } G_r \in \mmbl \text{ are disjoint} \bigg\},
\end{split}
\end{equation*}
where $\mathds{1}_G$ is the indicator function. Now let
\begin{align*}
\phi_1 (t) = \sum^M_{i=1} a_i \mathds{1}_{D_i} (t) \leq T(t)
\end{align*}
be any such function on $\mathcal{S}$. Assume wlog that $\phi_1$ is a simple function in canonical form, that is, $a_i \in [0,1], D_i \in \mmbl, D_i \neq \emptyset$ and $D_i$ form a partition of $\mathcal S$. In the rest of the proof, we construct a second simple function
\begin{align*}
\phi_2 (t) = \sum^N_{j=1} b_j \mathds{1}_{E_j} (t) \leq T(t)
\end{align*}
(where $b_j \in [0,1], E_j \in \mmbl, E_j$ are disjoint) such that
\begin{align*}
\sum^N_{j=1} b_j \mker(s_2, E_j ) \geq \left( \sum^M_{i=1} a_i \mker(s_1, D_i ) \right) - \left(1-\left(1-\varepsilon\right)^k\right)
\end{align*}
in order to show $\prob{k}{s_2}{\mathrm{TRACE}} \geq \prob{k}{s_1}{\mathrm{TRACE}} - \left(1-\left(1-\varepsilon\right)^k\right)$.
\\\\
Take $\{E_1,\dots,E_N\}$ to be the (measurable) partition of $\mathcal S$ generated by the equivalence relation induced by belonging to exactly the same $R_\varepsilon(D_i)$, so that $N \leq 2^M$. That is, $\forall x,y \in \mathcal S, \, x$ and $y$ are in the same $E_j$ iff we have: $\forall i,\, x \in R_\varepsilon(D_i) \iff y \in R_\varepsilon(D_i)$.

We now use Theorem \ref{them:Bollobas} to find appropriate $b_j$. Denote by $\mathcal{B}(I)$ the Borel $\sigma$-algebra on $I$ for any interval $I \in \mathbb{R}$. Denote by $l$ the Borel measure on $\mathbb{R}$. Let $*$ be some fresh element not in $\mathcal S$ (which we will use as padding to represent the $\varepsilon$ error), and let $\Upsilon \coloneqq  \{*\} \times [0,\varepsilon]$.
With reference to Theorem \ref{them:Bollobas}, take:

\begin{itemize}
\item $Z \coloneqq \left( \mathcal{S} \times [0,1] \right) \uplus \Upsilon$
\item $\mathcal F$ to be the smallest $\sigma$-algebra on $Z$ containing $\left( \mmbl \times \mathcal{B}([0,1]) \right) \,\uplus\, \left( \{*\} \times \mathcal{B}([0,\varepsilon])\right)$ 
\item $\mu$ to be the unique measure on $\mathcal F$ satisfying: \begin{description}
\item for $S \times I \; \in \; \mmbl \times \mathcal{B}([0,1]), \;\; \mu(S \times I) = \mker(s_2, S) \times l(I)$
\item for $\{*\} \times I \; \in \; \{*\} \times \mathcal{B}([0,\varepsilon]), \;\; \mu(\{*\} \times I) = l(I)$ \end{description}
\item $A \coloneqq \{1,\dots,M\}$
\item $X_i \coloneqq \left( R_\varepsilon (D_i) \times [0,1]\right) \uplus \Upsilon$
\item $\lambda_i \coloneqq \mker(s_1,D_i)$. 
\end{itemize}

Now, $Z$ is atomless since real intervals with the Borel measure are atomless. $X_i$ have finite measure, since $\mu(Z) = 1 + \varepsilon$. Since $s_1 R_\varepsilon s_2$, we have $\mker(s_2,R_\varepsilon (S)) + \varepsilon \geq \mker(s_1,S) \; \forall S \in \mmbl$. Therefore, we have that for any (finite) $B \subseteq A$,
\begin{align}
  \begin{aligned}
    \mu \left( \cup_{i \in B} X_i \right)
    &= \mu \left( \cup_{i \in B} {R_\varepsilon (D_i)} \times [0,1] \right) + \mu\left( \{*\} \times [0,\varepsilon]\right)
    \\&=\mker \left(s_2, \cup_{i \in B} {R_\varepsilon (D_i)} \right) + \varepsilon
    \\&=\mker \left(s_2, R_\varepsilon ( \cup_{i \in B}D_i) \right) + \varepsilon
    \\&\geq \mker \left(s_1, \cup_{i \in B}D_i \right)
    \\&=\sum_{i \in B} \lambda_i. 
  \end{aligned}
  \nonumber
\end{align}
Therefore by Theorem \ref{them:Bollobas}, there exist $Y_i \in \mathcal F$ s.t.
\begin{align*}
  &Y_i \subseteq X_i
\\&\mu(Y_i) = \mker(s_1,D_i)
\\&\mu\left(Y_i \cap Y_{i'}\right) = 0 \;\; \forall i \neq i'. 
\end{align*}
Let $E'_j \coloneqq E_j \times [0,1]$ so that $\mu(E'_j) = \mker(s_2,E_j)$. Take
\begin{align*}
b_j \coloneqq 
\begin{cases}
  \sum^M_{i=1} \frac{\mu(Y_i \cap E'_j)}{\mu(E'_j)} (a_i-\delta)^+ & \text{if}\ \mu(E'_j) > 0 \\
  0 & \text{otherwise}
\end{cases}
\end{align*}
where $(a_i-\delta)^+ \coloneqq \max\{0,a_i-\delta\}$. The $b_j$ here serve the same purpose as the coefficients in Equations (\ref{eqn:break up and match transitions1}) and (\ref{eqn:break up and match transitions2}) in the sketch proof of this theorem in Section \ref{sect: approx bisim implies approx trace equiv}. 

We now show $\phi_2(t) \leq T(t)$.
Fix any $j\in\{1,\dots,N\}$ s.t. $\mu(E'_j)>0$, and fix any $t \in E_j$.
%\begin{align*}
%b_j = \sum_{i \mid Y_i \cap E'_j \neq \emptyset} \left( a_i - \delta \right) \frac{\mu(Y_i \cap E'_j)}{\mu(E'_j)}. 
%\end{align*}
Let $i$ be such that $Y_i\cap E'_j \neq \emptyset$. $Y_i \subseteq X_i$, so $R_\varepsilon(D_i) \cap E_j \neq \emptyset$, so $E_j \subseteq R_\varepsilon(D_i)$ by construction of $E_j$, so $t \in R_\varepsilon(D_i)$. There exists $x \in D_i$ such that $x R_\varepsilon t$. We have $a_i = \phi_1(x) \leq T(x)$, and so by the induction hypothesis we have
\begin{align*}
T(t) \geq T(x) - \delta \geq a_i - \delta. 
\end{align*}
This holds for every $i$ such that $Y_i \cap E'_j \neq \emptyset$, and $\sum_{i \mid Y_i \cap E'_j \neq \emptyset} \frac{\mu(Y_i \cap E'_j)}{\mu(E'_j)} \leq 1$, so we get
\begin{align*}
T(t)
\geq \sum_{i \mid Y_i \cap E'_j \neq \emptyset} \frac{\mu(Y_i \cap E'_j)}{\mu(E'_j)} (a_i-\delta)^+
= b_j
= \phi_2(t)
\end{align*}
This holds for any $t \in E_j$ where $\mu(E_j)>0$. We also have $T(t) \geq 0 = \phi_2(t)$ for $t \in E_j$ where $\mu(E_j)=0$. Hence for any $t \in \mathcal{S}$, we have $T(t) \geq \phi_2(t)$.
\\\\
We now show that 
\begin{align*}
\sum^M_{i=1} a_i \mker(s_1, D_i ) - \sum^N_{j=1} b_j \mker(s_2, E_j ) \leq 1-\left(1-\varepsilon\right)^k
\end{align*}
Noting that $E'_j$ form a partition of $\mathcal S \times [0,1]$, we have
\begin{align}
  \begin{aligned}
    \sum^N_{j=1} b_j \mker(s_2, E_j )
    &= \sum_{j \mid \mu(E_j)>0} \mker(s_2, E_j ) \sum^M_{i=1} \frac{\mu(Y_i \cap E'_j)}{\mu(E'_j)} \left( a_i - \delta \right)^+
    \\&= \sum^N_{j=1} \sum^M_{i=1} \mu(Y_i \cap E'_j) \left( a_i - \delta \right)^+
    \\&= \sum^M_{i=1} \left( a_i - \delta \right)^+ \sum^N_{j=1} \mu(Y_i \cap E'_j)
    \\&= \sum^M_{i=1} \left( a_i - \delta \right)^+ \mu \left( \bigcup^N_{j=1} \left(Y_i \cap E'_j\right)\right)
    \\&= \sum^M_{i=1} \left( a_i - \delta \right)^+ \mu( Y_i \setminus \Upsilon). 
  \end{aligned}
  \nonumber
\end{align}
We obtain  
\begin{align*}
    \sum^M_{i=1} a_i \mker(s_1, D_i ) - \sum^N_{j=1} b_j \mker(s_2, E_j )
    =\sum^M_{i=1} a_i \mker(s_1, D_i ) - \sum^M_{i=1} \left( a_i - \delta \right)^+ \mu( Y_i \setminus \Upsilon). 
\end{align*}
We now apply Lemma \ref{lemma:bound algebra} to the RHS of this equation. We check the conditions of Lemma \ref{lemma:bound algebra} (in fact, by taking $y_i \coloneqq (a_i-\delta)^+$, we get $y'_i=y_i$ in the proof of Lemma \ref{lemma:bound algebra}):
\begin{align*}
  &a_i-(a_i-\delta)^+ \leq \delta
\\&\sum^M_{i=1} \mker(s_1,D_i) \leq 1
\\&\sum_{i \mid \mu\left(Y_i \setminus \Upsilon \right) < \mker(s_1,D_i)} \left( \mker(s_1,D_i) - \mu\left(Y_i \setminus \Upsilon \right) \right) \leq \varepsilon, 
\end{align*}
where the last condition holds because $\forall i \neq i' \;\; \mu\left(Y_i \cap Y_{i'}\right) = 0$, and so we have for any $J \subseteq \{1, \dots, M \}$,
\begin{align*}
\sum_{i \in J}{\mu\left(Y_i \setminus \Upsilon \right)}
&= \mu\left(\bigcup_{i \in J}\left(Y_i \setminus \Upsilon \right)\right)
\\&= \mu\left(\bigcup_{i \in J} Y_i\right) - \mu \left( \Upsilon \cap \bigcup_{i \in J} Y_i \right)
\\&\geq \mu\left(\bigcup_{i \in J} Y_i\right) - \varepsilon
\\&= \sum_{i \in J} {\mker(s_1,D_i)} - \varepsilon. 
\end{align*}
So, by Lemma \ref{lemma:bound algebra},
\begin{align*}
\sum^M_{i=1} a_i \mker(s_1, D_i ) - \sum^N_{j=1} b_j \mker(s_2, E_j )
&\leq \varepsilon + \delta - \varepsilon\delta
\\&= 1-\left(1-\varepsilon\right)^k. 
\end{align*}
In conclusion, $\phi_2$ has all of the required properties.
%\qed

\section{Proof of Theorem \ref{prop:partition for approx bisim}} 

\begin{proof}
Wlog, assume $\msspaceC$ and $\msspaceA$ are disjoint. We will prove that $\sC, \sA$ are $\varepsilon$-bisimilar as states in the direct sum LMC $\mchainU = \mtupleU$, as per Definition \ref{def:direct sum of LMCs}.
\\\\\noindent
Let
$$
R_\varepsilon \coloneqq \Big\{ (\sA_k, \sA_k), (\sA_k, \sC), (\sC, \sA_k), (\sC, {\sC}')
\Bigm| k \in \{ 1, \cdots, N \},\;\; \sC,{\sC}'\!\! \in P_k \Big\}
$$
be a binary relation over $\msspaceU$. Here, we have taken $R_\varepsilon$ to be the largest $\varepsilon$-bisimulation relation over $\mchainU$; other choices of $R_\varepsilon$ also work. Since the other conditions are clearly satisfied, to prove that $R_\varepsilon$ is an $\varepsilon$-bisimulation relation we will only check
\begin{align*}
\mkerU \left( s_1, R_\varepsilon(T) \right) \geq \mkerU \left( s_1, R_\varepsilon(T) \right) - \varepsilon
\end{align*}
for all $s_1,s_2 \in \msspaceU$, $T \in \mmblU$.
\\\\\noindent
Decompose $T$ as 
$T = V \uplus \{ \sA_i \mid i \in I\}$, 
where $V = T \cap \msspaceC$, $\{ \sA_i \mid i \in I\} = T \cap \msspaceA$, $I \subseteq \{ 1, \cdots, N \}$, and let $J \coloneqq \big\{ j \in \{1, \cdots, N\} \bigm| T \cap P_j \neq \emptyset \big\}$, so that
\begin{align*}
R_\varepsilon(T) = \{\sA_i \mid i \in I \cup J \} \uplus \bigcup_{j \in I \cup J} P_j. 
\end{align*}
Let $(s_1,s_2) \in R_\varepsilon$.
For $(s_1,s_2) = (\sA_k,\sA_k)$ where $k \in \{1, \cdots, N\}$,
\begin{align*}
\mkerU(s_1, R_\varepsilon(T))
   &= \mkerA(s_1, R_\varepsilon(T) \cap \msspaceA)
\\ &\geq \mkerA(s_1, \{\sA_i \mid i \in I \})
\\ &= \mkerA(s_2, \{\sA_i \mid i \in I \}) \indent\text{(since $s_1 = s_2$)}
\\ &= \mkerA(s_2, T \cap \msspaceA)
\\ &= \mkerU(s_2, T). 
\end{align*}
For $(s_1,s_2) = (\sC,{\sC}')$ where $\sC,{\sC}' \in P_k$,
\begin{align*}
\mkerU(s_1, R_\varepsilon(T))
   &= \mkerC(s_1, R_\varepsilon(T) \cap \msspaceC)
\\ &\geq \mkerC(s_1, \cup_{j \in J} P_j)
\\ &\geq \mkerC(s_2, \cup_{j \in J} P_j) - \varepsilon
\\ &\indent \text{(by assumption, since $s_1,s_2$ are in the same partition $P_k$)}
\\ &\geq \mkerC(s_2, T \cap \msspaceC) - \varepsilon
\\ &= \mkerU(s_2, T) - \varepsilon. 
\end{align*}
For $(s_1,s_2) = (\sA_k,\sC)$ where $k \in \{1, \cdots, N\}$, $\sC \in P_k$,
\begin{align*}
\mkerU(s_1, R_\varepsilon(T))
   &\geq \mkerA(s_1, \{\sA_j \mid j \in J \})
\\ &= \mkerC(\sC_k, \cup_{j \in J} P_j) \indent \text{(by def of $\mkerA$)}
\\ &\geq \mkerC(s_2, \cup_{j \in J} P_j) - \varepsilon
\\ &\indent \text{(by assumption, since $\sC,s_2$ are in the same partition $P_k$)}
\\ &\geq \mkerU(s_2, T) - \varepsilon. 
\end{align*}
For $(s_1,s_2) = (\sC,\sA_k)$ where $k \in \{1, \cdots, N\}$, $\sC \in P_k$, similarly we have
\begin{align*}
\mkerU(s_1, R_\varepsilon(T))
   &\geq \mkerC(s_1, \cup_{i \in I} P_i)
\\ &\geq \mkerC(\sC_k, \cup_{i \in I} P_i) - \varepsilon
\\ &= \mkerA(s_2, \{\sA_i \mid i \in I \}) - \varepsilon
\\ &= \mkerU(s_2, T). 
\end{align*} 
\qed
\end{proof}

\pagebreak
\section{Proof of Theorem \ref{prop:smooth kernel}}

\begin{proof}
\begin{align*}
\left| \mkerC(s_1, T) - \mkerC(s_2, T) \right|
\leq K \cdot \Vert s_1 - s_2 \Vert \cdot \lambda(T)
\leq K \cdot \delta(T) \cdot \lambda(T),
\end{align*}
and so
\begin{align*}
& \max_{J \subseteq \{1, \cdots, N\}} \left| \mkerC \left( s_1, \bigcup_{j \in J} P_j \right) - \mkerC \left( s_2, \bigcup_{j \in J} P_j \right) \right|
\\&\indent = \frac{1}{2} \sum^{N}_{j=1} \left| \mkerC(s_1, P_j) - \mkerC(s_2, P_j) \right|
\\&\indent \leq \frac{1}{2} K \cdot \max_{j \in \{1, \cdots, N \}} \delta(P_j) \cdot \lambda(\msspaceC)
\\&\indent \leq \varepsilon. 
\end{align*}
\qed
\end{proof}

\section{Proof of Theorem \ref{thm:casestudy}}

\begin{proof}
For any $r\in\{0,1\}$, $h\in\{0,\cdots,N-1\}$ and $s_1,s_2 \in P_{r,h}$, assume wlog $s_1 = (r, x_1), s_2 = (r,x_2)$ and $x_1 \leq x_2$. Then, $\mkerC (s_1, \cdot), \mkerC (s_2, \cdot)$ differ only on $V \subseteq \msspaceC$, where
\begin{align*}
V = \begin{cases}
  \{0,1\} \times \left[ \frac{1+x_1}{2}, \frac{1+x_2}{2} \right) & \text{if}\ r=1 \\
  \{0,1\} \times \left[ \frac{x_1}{2}, \frac{x_2}{2} \right) & \text{if}\ r=0. 
\end{cases}
\end{align*}
Therefore, we have for all $T \in \mmblC$,
\begin{align*}
& \left| \mkerC (s_1, T) - \mkerC (s_2, T) \right|
\\&\indent\indent 
\leq \sup_{s\in\msspaceC} \mkerC(s,V)
\\&\indent\indent 
\leq \sup_{p_\mathrm{rain} \in \left[0,1\right]} \Big| p_\mathrm{rain} (x_2-x_1)+ (1-p_\mathrm{rain}) (x_2-x_1) \Big|
\\&\indent\indent 
= \nicefrac{1}{N}. 
\end{align*}
\qed
\end{proof}

\section{Case study - model checking abstract model $\mchainA$ with PRISM}
\label{Appendix: PRISM code}

$\mchainA$ from Section \ref{sect:application case study} is model checked algorithmically by setting up the PRISM \cite{KNP11} model in Figure \ref{fig:PRISM}.

\begin{figure}[H]
\caption{\textbf{PRISM model for $\mchainA$ with $N=1000$.} 
The length of this textual specification grows linearly with $N$, and it was generated with the help of a script.}
\label{fig:PRISM}
\begin{framed}
\begin{lstlisting}
// PRISM specification for the Abstract Model.

dtmc
formula N = 1000;
formula pToRain = r = 1 ? 1/4 + 3/4 * h/N : 3/4 * h/N;
module weatherAbstractModel

// State space
r : [0..1];
h : [0.. (N-1)];

[] true ->
   (pToRain * 2/(N + h) * max(min((N+h)/2 - 000, 1), 0))
        : (r'=1) & (h'=0)
 + (pToRain * 2/(N + h) * max(min((N+h)/2 - 001, 1), 0))
        : (r'=1) & (h'=1)
<<< ... TRIMMED (same pattern for 997 lines) ... >>>
 + (pToRain * 2/(N + h) * max(min((N+h)/2 - 999, 1), 0))
        : (r'=1) & (h'=999)
 + ((1-pToRain) * 2/(2*N-h) * max(min(000 + 1 - h/2, 1), 0))
        : (r'=0) & (h'=0)
 + ((1-pToRain) * 2/(2*N-h) * max(min(001 + 1 - h/2, 1), 0))
        : (r'=0) & (h'=1)
<<< ... TRIMMED (same pattern for 997 lines) ... >>>
 + ((1-pToRain) * 2/(2*N-h) * max(min(999 + 1 - h/2, 1), 0))
        : (r'=0) & (h'=999);

endmodule
\end{lstlisting}
\end{framed}
\end{figure}

The probability of there being 2 consecutive days of rain over the next three days is computed by expressing the PRISM property  
\begin{lstlisting}[basicstyle=\footnotesize]
filter(state, P=? [
  ( (X (r=1)) & (X X (r=1)) )
  | ( (X X (r=1)) & (X X X (r=1)) )
], r=0&h=500)
\end{lstlisting}

On a 2.5 GHZ processor using PRISM 4.3, it took 24 sec to parse and build the model,
and 1 sec to compute this particular probability.

\end{document}